\documentclass[useAMS,usenatbib]{mn2e}
\usepackage{url,ulem,times,graphicx,amsmath,amsfonts,amssymb,color,epsfig,epstopdf}
\newcommand{\beq}{\begin{eqnarray}}  
\newcommand{\eeq}{\end{eqnarray}}

\newcommand{\hMpc}{{\ifmmode{h^{-1}{\rm Mpc}}\else{$h^{-1}$Mpc }\fi}}  
\newcommand{\hGpc}{{\ifmmode{h^{-1}{\rm Gpc}}\else{$h^{-1}$Gpc }\fi}}  
\newcommand{\hmpc}{{\ifmmode{h^{-1}{\rm Mpc}}\else{$h^{-1}$Mpc }\fi}}  
\newcommand{\hkpc}{{\ifmmode{h^{-1}{\rm kpc}}\else{$h^{-1}$kpc }\fi}}  
\newcommand{\hMsun}{{\ifmmode{h^{-1}{\rm {M_{\odot}}}}\else{$h^{-1}{\rm{M_{\odot}}}$}\fi}}  
\newcommand{\hmsun}{{\ifmmode{h^{-1}{\rm {M_{\odot}}}}\else{$h^{-1}{\rm{M_{\odot}}}$}\fi}}  
\newcommand{\Msun}{{\ifmmode{{\rm {M_{\odot}}}}\else{${\rm{M_{\odot}}}$}\fi}}  
\newcommand{\msun}{{\ifmmode{{\rm {M_{\odot}}}}\else{${\rm{M_{\odot}}}$}\fi}}  
\newcommand{\LCDM}{$\Lambda$CDM}

\begin{document}  

\title[The cosmic web]{A kinematic classification of the cosmic web} 
\author[Hoffman et al.] 
{\parbox[t]\textwidth{ Yehuda Hoffman$^1$, Ofer Metuki$^1$,  Gustavo Yepes$^2$,
  Stefan Gottl\"ober$^3$, Jaime E. Forero-Romero$^{3}$,  Noam I. Libeskind$^{3}$   \& Alexander Knebe$^2$} 
\vspace*{6pt} \\ 
 $^{1}$Racah Institute of Physics, Hebrew University, Jerusalem 91904, 
 Israel\\ 
 $^{2}$Grupo de Astrof\'{\i}sica, Universidad Aut\'onoma de Madrid, 
 Madrid E-28049, Spain \\ 
 $^{3}$Leibniz-Institut f\"ur Astrophysik Potsdam (AIP), An der Sternwarte 16, D-14482 
 Potsdam, Germany\\  
} 
\date{\today}  

\maketitle

\begin{abstract}  
A new approach for the classification of the cosmic web is presented. In extension of the previous work of \citet{2007MNRAS.375..489H} and  \citet{2009MNRAS.396.1815F} the new algorithm is based on the analysis of the velocity shear tensor rather than the gravitational tidal tensor. The procedure consists of the construction of the the shear tensor at each (grid) point in space and the evaluation of its three eigenvectors. A given point is classified to be either a void, sheet, filament or a knot according to the number of eigenvalues above a certain threshold, 0, 1 , 2, or 3 respectively. The threshold is treated as a free parameter that defines the web. 
The algorithm has been applied to a  dark matter only 
 simulation  of a box of  side-length $64\hmpc$ and $N=1024^3$ particles  with the framework of the WMAP5/$\Lambda$CDM model. The resulting velocity based cosmic web resolves structures down to $\lesssim 0.1\hmpc$ scales, as opposed to the $\approx 1 \hmpc$ scale of the tidal based web. The under-dense regions are made of extended voids bisected by planar sheets, whose density is also below the mean. The over-dense regions are vastly dominated by the linear filaments and knots. The resolution achieved by the velocity based cosmic web  provides a platform for studying the formation of halos and galaxies within the framework of the cosmic web.

\end{abstract}

\section{Introduction}  
\label{sec:intro}

The visual appearance of the large scale structure (LSS) of the universe is of a web, the so called cosmic web \citep{1996Natur.380..603B}. The large scale distribution of galaxies in the observed universe as well as the distribution of the dark matter (DM) as inferred from its gravitational lensing  and reconstructions from large galaxy surveys  give the appearance of mass and light distributed in a web-like structure dominated by linear filaments and concentrated compact knots, thereby leaving behind vast extended regions of no or a few galaxies and of low density
 \citep{2009MNRAS.400..183K,2010MNRAS.409..355J,2011arXiv1107.1062M,2011arXiv1108.1008W}. 
Direct mapping of the mass distribution by weak lensing reveals a time evolving  loose network of filaments, which connects rich clusters of galaxies \citep{2007Natur.445..286M}. The extreme low resolution of the weak lensing maps cannot reveal the full intricacy of the cosmic web, and in particular the difference between filaments and sheets, yet they reveal a web structure that serves as a  gravitational scaffold into which gas can accumulate, and stars can be built.
This poses an intriguing challenge of the mathematical classification and quantification of   the cosmic web. The motivation for such an endeavor is twofold. On the one hand the cosmic web is there and so we want to describe it mathematically. On the other hand the web classification might provide some extra parameters that quantify the environment within which DM halos and galaxies form.
The properties of galaxies are observed to depend on their environment \citep{1980ApJ...236..351D,2005ApJ...629..143B,2005ApJ...634...51A,2005MNRAS.363L..66G,2007ApJ...654...53M,2011arXiv1107.0017F}.
Further more the dynamics of sub-halos and satellite galaxies suggests a possible dependence on the environment within which their parent halos reside  \citep{2004ApJ...603....7K,2005MNRAS.363..146L,2011MNRAS.411.1525L}.
A web classification might extend our theoretical tools for understanding such a dependance.

Translating the visual impression of a cosmic web into a mathematical formulation is not  trivial. 
The  thorough   review  of \cite{2008MNRAS.387..933C}  of void finders    provides some perspective to the  general issue of web classifiers. 
This has been pursued along two different lines, the geometric and the dynamic approaches. 
The geometric strategy focuses on the point process exhibited   by the distribution of galaxies   or DM 
halos in simulations, say,  and describes it mathematically. This has been often applied independently of any dynamical context \citep{1999MNRAS.302..111L,2006MNRAS.366.1201N,2007A&A...474..315A,2008MNRAS.383.1655S}.
The dynamic approach   has its roots in the seminal work of \cite{1970A&A.....5...84Z}, which led to the 'Russian school of structure formation' (e.g. \cite{1982GApFD..20..111A,1983MNRAS.204..891K})
 The   Zeldovich approximation has been the first analytical tool that enables the tracking of the formation of aspherical objects, thereby accounting for the formation of voids, sheets, filament and knots. A recent application of that approach to the cosmic web is presented in 
\citet{2008ApJ...688...78L}

The geometric approach uses objects (galaxies or DM halos) which are per definition at a position of locally high density. \citet{2007MNRAS.375..489H} proposed a very attractive dynamic classification of the cosmic web. They presented an algorithm by which each point in space can be classified  as   being either a void, sheet, filament or a knot point.  Namely, an attribute  is assigned at each point in space much in the same way as it is characterized by its density and velocity. 
This attribute    depends, in turn,  on the the dynamics of the LSS, namely on the density and/or the velocity fields.

The   \citet{2007MNRAS.375..489H} dynamic web classification   is based on counting the number of positive eigenvalues of the tidal tensor, i.e. the Hessian of the gravitational potential. The number ranges over $0, 1, 2$ and $3$, corresponding  to a void, sheet, filament and a knot attribute. This web classifier is of a dynamic nature and it is closely related to the equations of motion that dictate the dynamics of the
 growth of structure. One can consider the web classifier to be a dynamical field that  characterizes the large scale structure of the universe, a field that accepts only four possible numbers. 
This approach is driven by the Zeldovich approximation  and its web classification agrees remarkably well with the visual impression one has in viewing the large scale structure (LSS) on linear and quasi-linear scales, i.e. scales larger than a few Megaparsecs. However, in going down to smaller scales the Hahn et al approach fails to recover the  fine web exhibited by high resolution  numerical simulations. A partial remedy to this shortcoming was provided by \cite{2009MNRAS.396.1815F} (FR09), who relaxed the assumption of the null  threshold which is used to classify the web. FR09 argues  that the value of the properly normalized  threshold should be around unity. Indeed the cosmic web defined by a non-zero threshold provides a much better description on the non-linear cosmic web, down to the Megaparsec scale.

Inspection of high resolution N-body simulations reveals a web-like structure on very small scales as well. In simulations aimed at resolving galactic scale halos the web is observed down to the  virial radius  of such halos \citep{2011MNRAS.411.1525L}. Studying such simulations we have tried to extend the approach of FR09 down to sub-Megaparsec scales, but have failed. Namely, the method has not been able to resolve the   fine web of these scales. 

In the linear regime of the gravitational instability the velocity and gravitational fields are essentially   identical, up to some scaling that depends on the cosmological parameters. Hence,   the web classification of \cite{2007MNRAS.375..489H} and FR09 can be reformulated in terms of the (velocity) shear tensor. It follows that in the linear regime the tidal tensor based cosmic web and the velocity tensor based web are identical (hereafter these are defined as  T-web and V-web, respectively). Going   to the fully non-linear regime we expect the  T- and V-web to depart. Here, we study the cosmic web as revealed by studying the analysis of the fully non-linear velocity field. The algorithm proposed here is tested against a high resolution DM-only N-body simulation.

A general remark is due here, before proceeding to the main body of the paper. By itself the cosmic web is an ill-defined entity. The transition from knots to filaments, sheets and voids seems to be smooth and gradual, yet visual impression strongly suggests that the LSS is correctly characterized by a cosmic web. This resonates  somewhat with the issue of DM halos. Again visual impression strongly suggests that the dynamics of gravitational instability leads to the formation of bound compact entities called DM halos, yet these objects have no strict boundaries and therefore their definition would rely on somewhat arbitrary chosen free parameter(s). However, even if the gravitational collapse in an expanding universe is still  not fully understood a lot of insight and understanding is provided by the spherical top-hat collapse model 
\citep{1972ApJ...176....1G, BT}. It follows that 
the majority of
halo finders are based on the model and the free parameter that defines a given halo finder, e.g. the mean density of a halo or the linking length that defines friends-of-friends halos, is derived from the top-hat model. 
No such model exists that can provide a quantitative  estimation of the value of one or more parameters that define the web. Hence, web classifier algorithms are bound to depend on some free parameters that cannot be determined from first principles. The web that emerges from the geometric web classifiers depends strongly on 
the type  or properties  of objects used to trace the web, such as
the mass of DM halos in simulations or the luminosity and spectral type of observed galaxies. The web that emerges from the dynamic web classification depends on the chosen smoothing, i.e resolution, and the value of the adopted threshold. No attempt is made here to optimize the values of these two parameters as the optimization depends on the specific problem one wishes to address with the V-web. Different problems end up with different 'best' parameters. We have chosen to use parameters that are best suited for the simulation studied here and the visual impression it leads to.

The paper consists of a description of the algorithm that classifies the cosmic web (\S \ref{sec:V-web}), a brief summary of the N-body simulation used here (\S \ref{sec:sim}) and a presentation of the results (\S \ref{sec:result} ). The paper concludes with a discussion of the V-web classification and its merits with respect to the analysis of structure and galaxy formation.

\section{Web Classification: algorithm}  
\label{sec:V-web}  

Consider a DM-only N-body simulation. A description of the simulation can be provided by the density and velocity fields evaluated on a finite grid, $\rho({\bf r})$ and ${\bf v}({\bf r})$. 
The T-web is defined by the eigenvalues of the Hessian of the gravitational potential which obeys the (rescaled) Poisson equation:
\begin{equation} 
\label{eq:poisson} 
\nabla^2 \phi = \Delta. 
\end{equation} 
Here $\phi$ is the gravitational potential rescaled by $4\pi G\bar{\rho}$, $\bar{\rho}$ is the mean cosmological density, and the density ($\rho$) is given by $\Delta=\rho / \bar{\rho}$.
The tidal tensor is defined as the Hessian of $\phi$, namely
 \begin{equation} 
\label{eq:hessian} 
T_{\alpha\beta}= {\partial ^2 \phi \over \partial r_\alpha \partial r_\beta},
\end{equation} 
with $\alpha, \beta = x, y $ and $z$.
The three eigenvalues of the tidal tensor are denoted here by $\lambda{^T_i}$, where $i=1, 2$ and $3$. (Throughout the paper  it   is assumed that the eigenvalues are arranged in decreasing order.) Note that 
\begin{equation} 
\label{eq:Delta} 
\Delta({\bf r}) = \lambda{^T_1}({\bf r}) +  \lambda{^T_2}({\bf r})  + \lambda{^T_3}({\bf r}). 
\end{equation} 
The V-web is defined in terms of the shear tensor, which is rescaled here and is written as:
\begin{equation}
\label{eq:Sigma}
\Sigma_{\alpha\beta} = -{1\over2} \big( {\partial  v_\alpha \over \partial r_\beta}  +  {\partial  v_\beta \over \partial r_\alpha}  \big) / H_0,
\end{equation}
where $H_0$ is the Hubble constant. The eigenvalues of $\Sigma_{\alpha\beta} $ are denoted here as $\lambda{^V_i}$ ($i=1, 2$ and $3$).

The T-web and V-web are classified here by the number of the eigenvalues above a threshold, $\lambda{^T_{\mathrm{th}}}$ and 
$\lambda{^V_{\mathrm{th}}}$.
We have not attempted   to determine the values of these thresholds but rather to treat them as free parameters which define the corresponding webs. Our guiding principle is to adopt values which best reproduce  the visual impression of the cosmic web.

\section{Simulation}  
\label{sec:sim}  

The web classification is applied here to a  simulation performed under the framework of the Constrained Local Universe Simulations
(CLUES) project\footnote{{\texttt http://www.clues-project.org/}},
whose aim is to perform cosmological simulations that reproduce the
local large scale structure in the Universe as accurately as current
observation admits. 
The simulation is performed within the \LCDM\  cosmology, with  cosmological parameters  consistent with a WMAP5
cosmology:  a  density parameter  $\Omega_{m}=0.28$, a cosmological constant
$\Omega_{\Lambda} = 0.72$, a dimensionless Hubble parameter  $h=0.70$,
a spectral index of primordial density perturbations $n=0.96$ and a normalization $\sigma_{8}=0.817$.
The constrained nature of the simulation is ignored here and it is treated as a typical realization of the \LCDM/WMAP5 cosmology.

This is a pure DM simulation performed by the 
Gadget2 code within a computational box   with side length
$L_{\mathrm{box}}=64\hmpc$ using $1024^3$ particles, 
corresponding  to a particle  mass of  $m_{p}=1.839\times 10^{7} \hmsun$.
The density and velocity fields are evaluated on a $256^3$ grid, using Clouds in Cells (CIC) 
interpolation. Two CIC constructions are used here, the full computational box (BOX64) and an inner box of side length $8\hmpc$ (BOX8) placed at the center of the computational box. 
The CIC interpolation of the velocity field is performed by  number weighting over the particles in adjacent cells. 
 Differentiation of the velocity field is performed in Fourier space, using FFT.
The CIC procedure  smoothes the field under consideration over two grid units, and therefore no aliasing is introduced by the FFT transform.

\section{Results}  
\label{sec:result}  

\begin{figure*}  
\begin{center}  
\includegraphics[width=6.8cm]{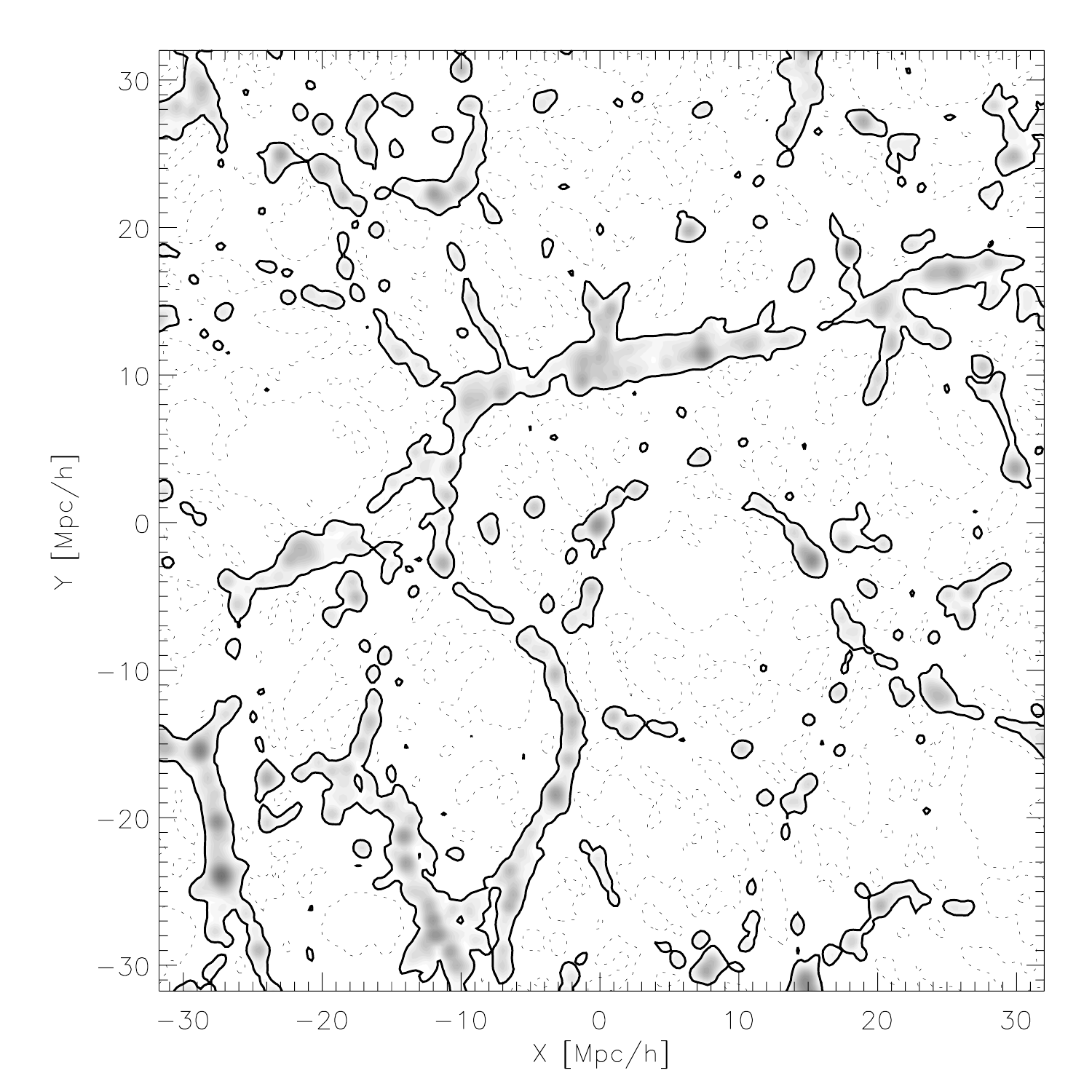}\hspace{0.5cm} 
\includegraphics[width=6.8cm]{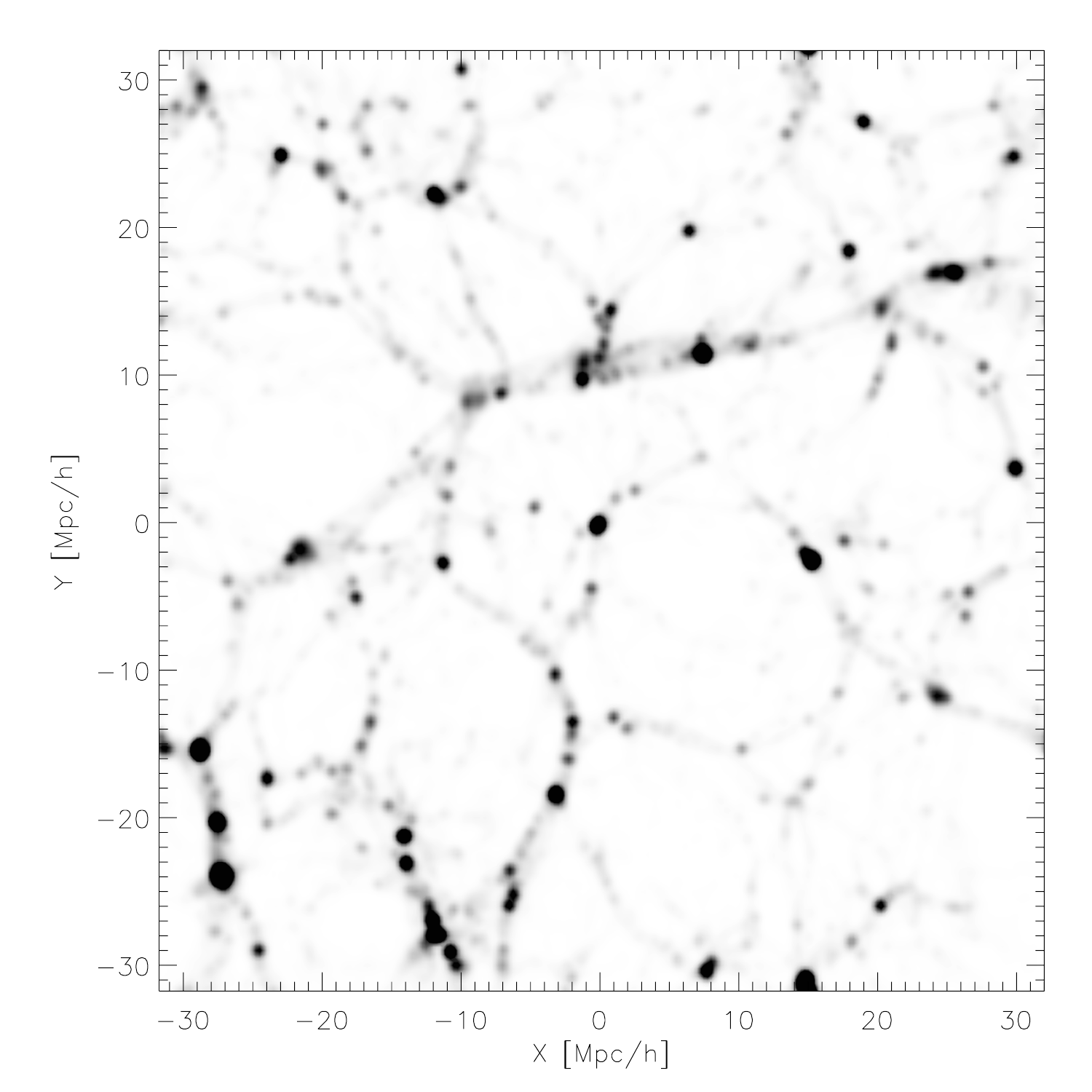}  
\includegraphics[width=6.8cm]{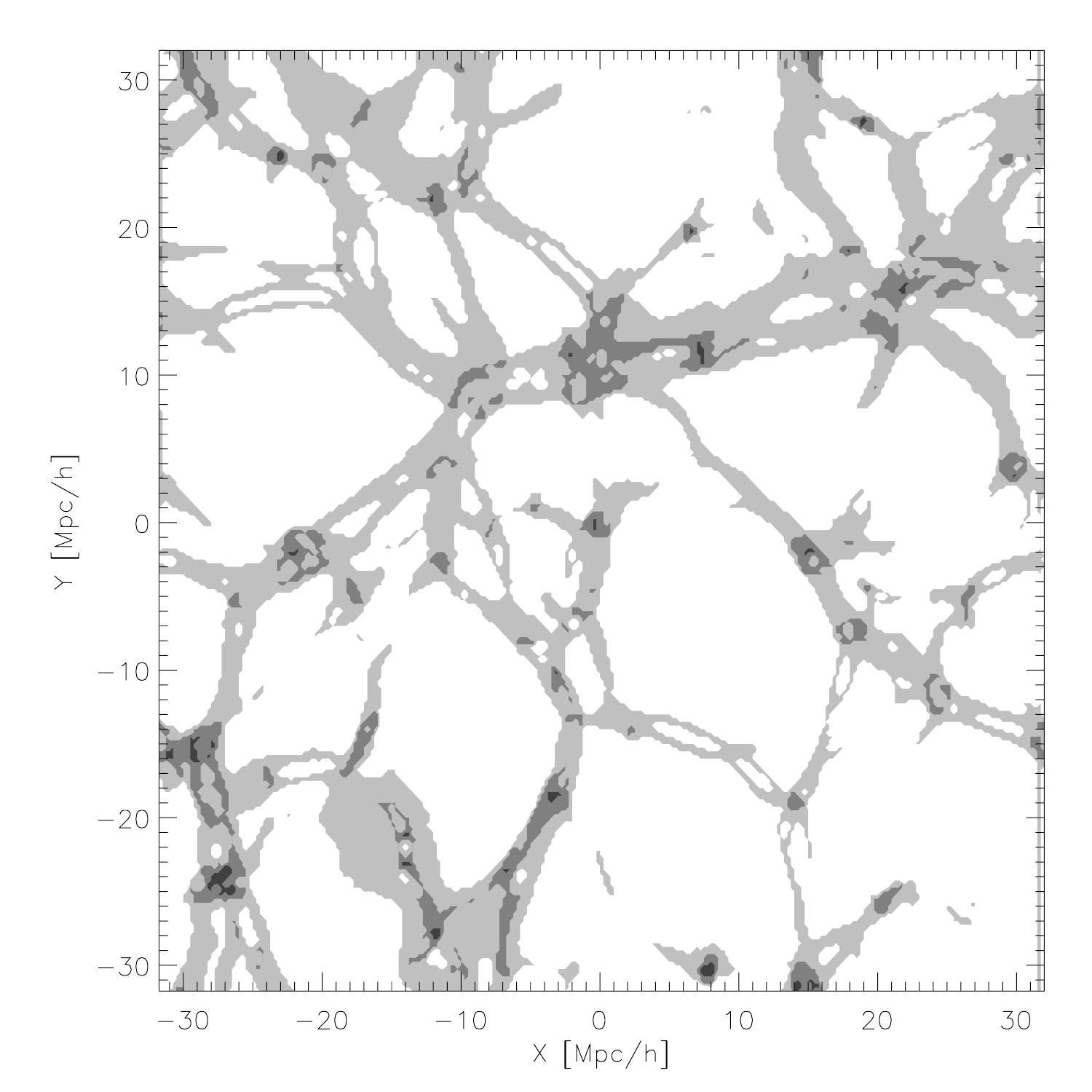}\hspace{0.5cm}  
\includegraphics[width=6.8cm]{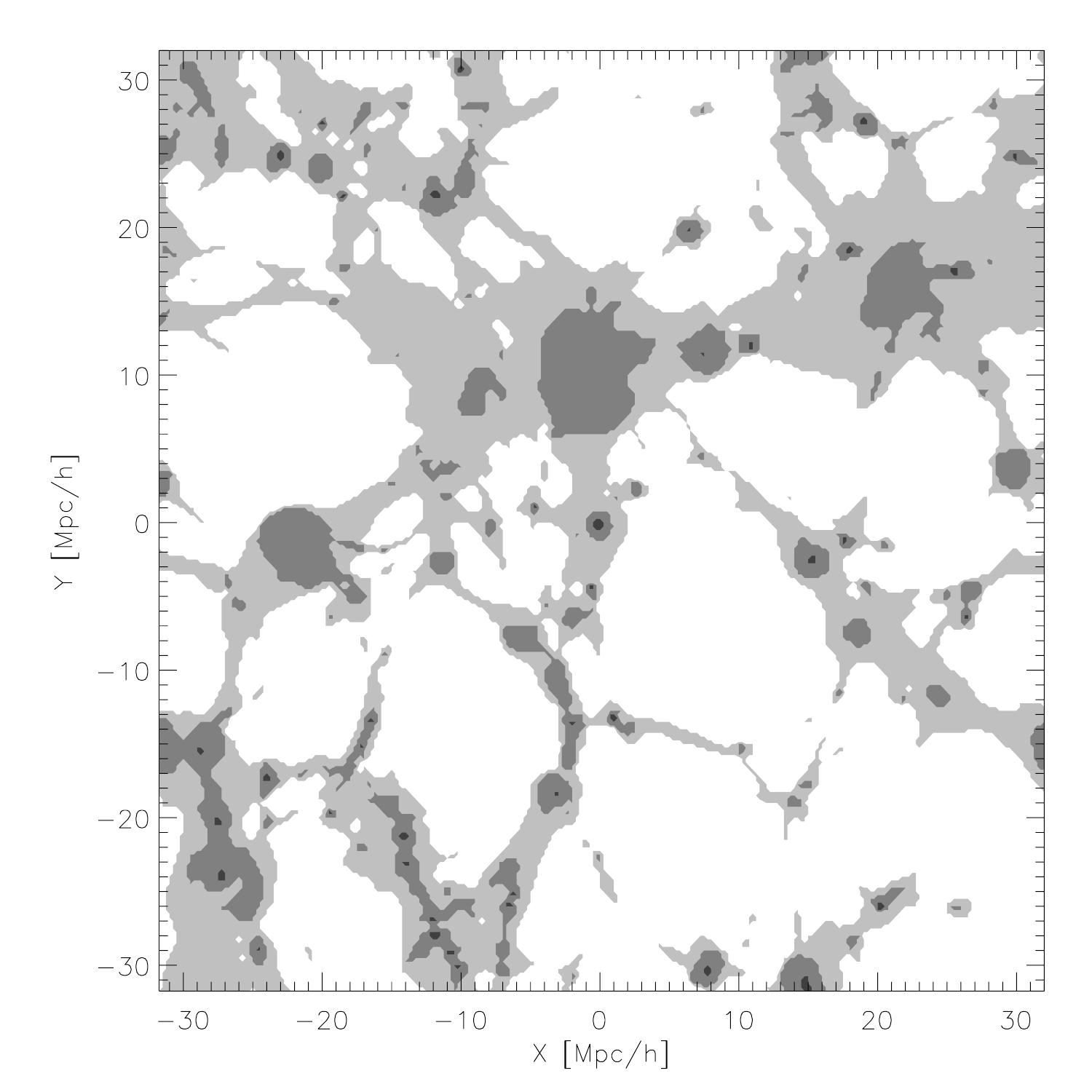}  
\includegraphics[width=6.8cm]{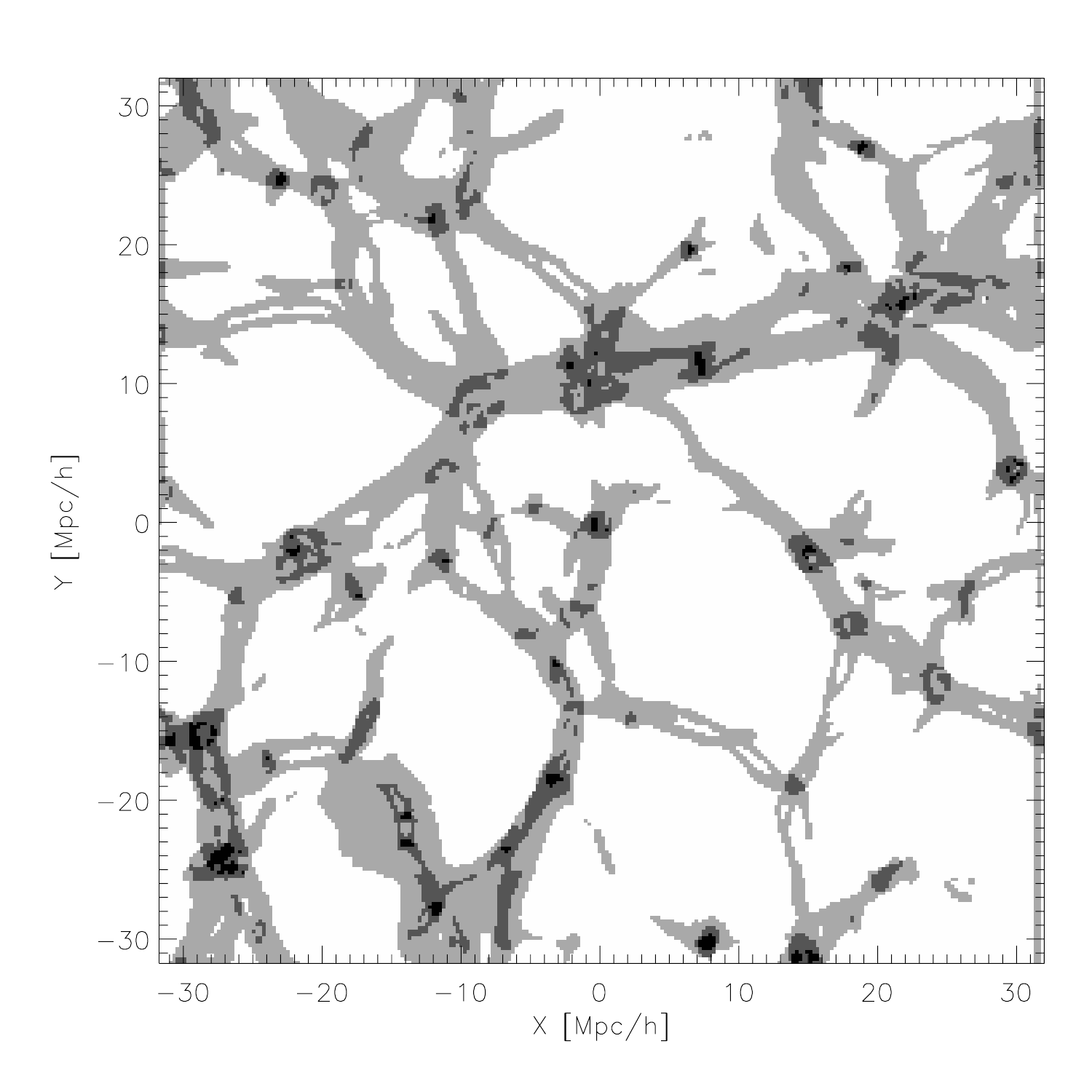}\hspace{0.5cm}  
\includegraphics[width=6.8cm]{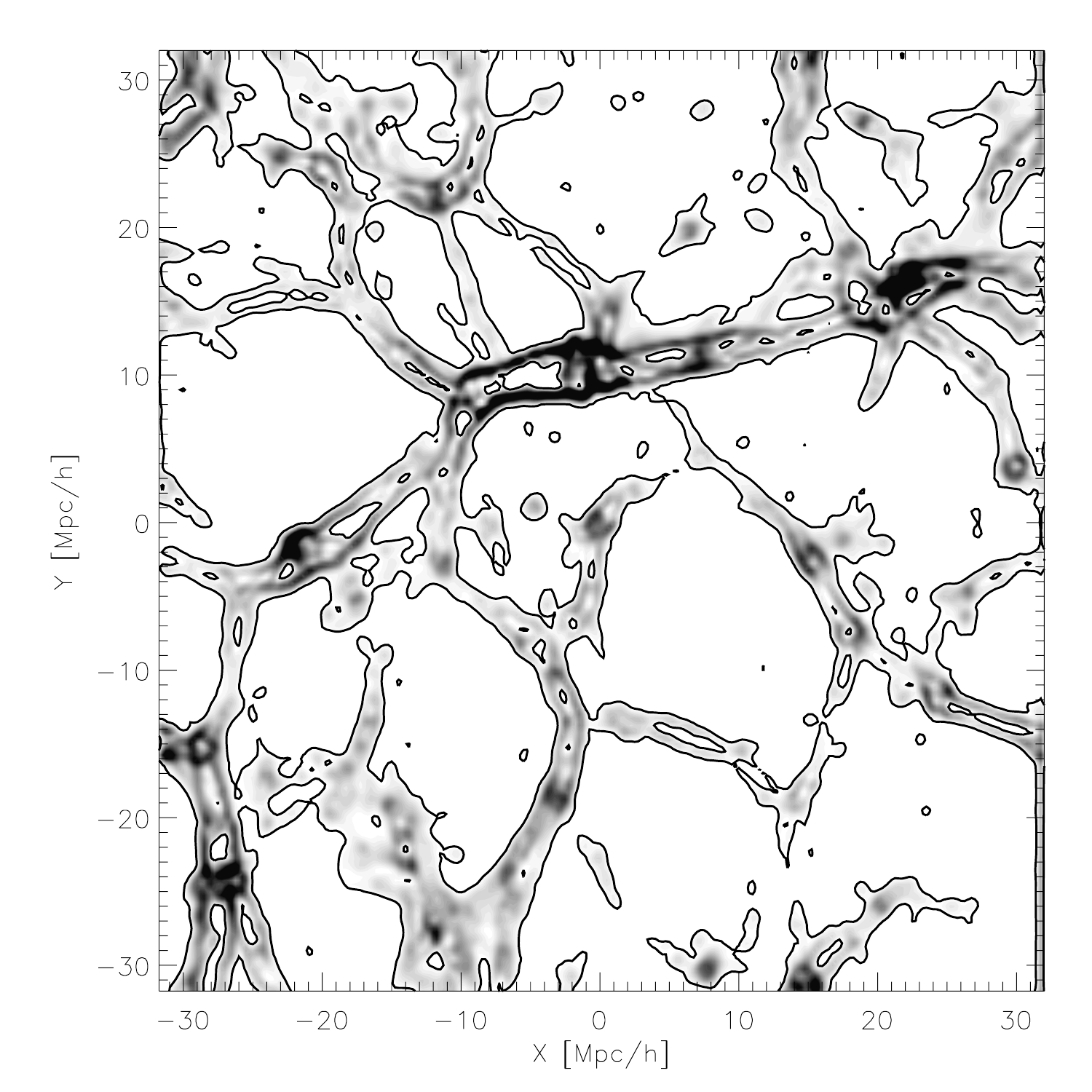}  
\caption{The normalized density   field and the cosmic web, 
based on the CIC density field of the full computational box (BOX64) spanned on a $256^3$ grid and Gaussian smoothed on the scale of $0.25 \hmpc$:
a.The density field presented by $\log\Delta$ (grey scale correspond to overdense  and dashed contours to under-dense regions.
 The solid contour represents the mean density. 
 (upper-left panel).
b. The linear density field (represented by grey scale color map; upper-right panel). 
c. The velocity based cosmic web generated with  $\lambda{^V_{th}}=0.44$ 
made of voids (white), sheets (light grey), filaments (dark grey) and knots (black). Note that the map presents a planar cut through the cosmic web, hence sheets appear as long  filaments and filaments as isolated compact regions  (middle-left panel).
d. The (gravitational) tidal   cosmic web generated with  $\lambda{^T_{th}}=0.7$
 (same color coding for the cosmic web, middle-right panel).
e. The multi-scale corrected V-web  (see Section \ref{sec:multi}, bottom-left panel).
f. The divergent of the velocity field: the grey scale shading  corresponds to $-\nabla\cdot\vec{v} \ge 0$ (bottom-right panel).
}  
\label{fig:box64-web} 
\end{center} 
\end{figure*}

Figure \ref{fig:box64-web} presents the analysis of the LSS delineated by one of the principal planes of the simulation. The analysis is applied to the BOX64 CIC field with a grid resolution of $0.25\hmpc$. Both the density and velocity fields have been Gaussian smoothed with a kernel of $R_s=0.25\hmpc$  so as to suppress numerical artifacts on the grid scale, 
in particular the unphysical preferred directions induced by the Cartesian grid.
The top panels show the  (scaled) density field, $\Delta({\bf r})$, in log (left) and linear (right) scales. The log scale map clearly distinguishes between the under- and over-dense regions. The middle panels show the V-web (left) and T-web (right). The threshold values used here are 
 $\lambda{^V_{\mathrm{th}}}=0.44$ 
 and $\lambda{^T_{\mathrm{th}}}=0.7$. 
 The threshold values of the two webs have been chosen so as to obtain the maximal spatial resolution and to provide the best visual match to the appearance of the density field.
 
 The grey scaling of the web maps is: white (voids), light grey (sheets), dark grey (filaments) and black (knots).
It should be noted that the maps shown here represent planar cuts,
 namely slices of one grid cell ($0.25\hmpc$) thickness,
 through a three dimensional structures, hence the sheets appear as roughly 1D filamentary objects while the filaments appear as isolated spots, reminding the appearance of the knots. 
The lower-right panel shows the (negative) of the divergence of the velocity fields, namely the sum of $\lambda{^V_i}$ at each grid point. 

Inspection of the different plots of Figure  \ref{fig:box64-web} reveals a rich cosmic web. 
A few immediate conclusions follow: 
\begin{enumerate}
  \item  The V-web provides a much superior description of the web structure that is revealed by the mass distribution, compared with the T-web. This is best exemplified by the dense object at $(X, Y, Z) \approx(-2,12,0)\hmpc$. The V-web reveals the delicate structure around the object while the T-web yields an unresolved blob over there. 
In fact, the T-web shown in the figure has been constructed with an optimal choice of the smoothing and threshold parameters, so as to achieve the maximum spatial resolution. In particular, a reduction in the Gaussian smoothing does not lead to an improvement in the resolution of the T-web.
  \item The web permeates through the under-dense regions, where large extended almost empty regions are bisected by sheets.  
  \item The sheets prevail mostly in the under-dense regions, compared with the filaments and knots which dominate the over-dense regions.
  \item Some of the sheets located in the under-dense regions exhibit a two caustics structure enclosing a void region. Namely, an  inner planar part has a diverging velocity field  (i.e., $\nabla \cdot {\bf v}  \ge 0$, is sandwiched between two caustic with a converging velocity field ($\nabla \cdot {\bf v}  < 0$). Such double sheets appear only in the under-dense regions. 
  \item The velocity divergence field follows very closely the density field, with almost a complete match of the under-dense    with the positive $-\nabla \cdot {\bf v}$ regions. Such a correspondence is the key for the success of the V-web in tracing the density field.
Note that in the linear theory of gravitational instability    under-dense points are always associated with a divergent flow. 
It follows that the  emergence of under-dense regions with converging flows is a pure 
non-linear dynamical effect. The present V-web algorithm classifies these objects as sheets.
\end{enumerate}

\begin{figure*}  
\begin{center} 
\includegraphics[width=5.8cm]{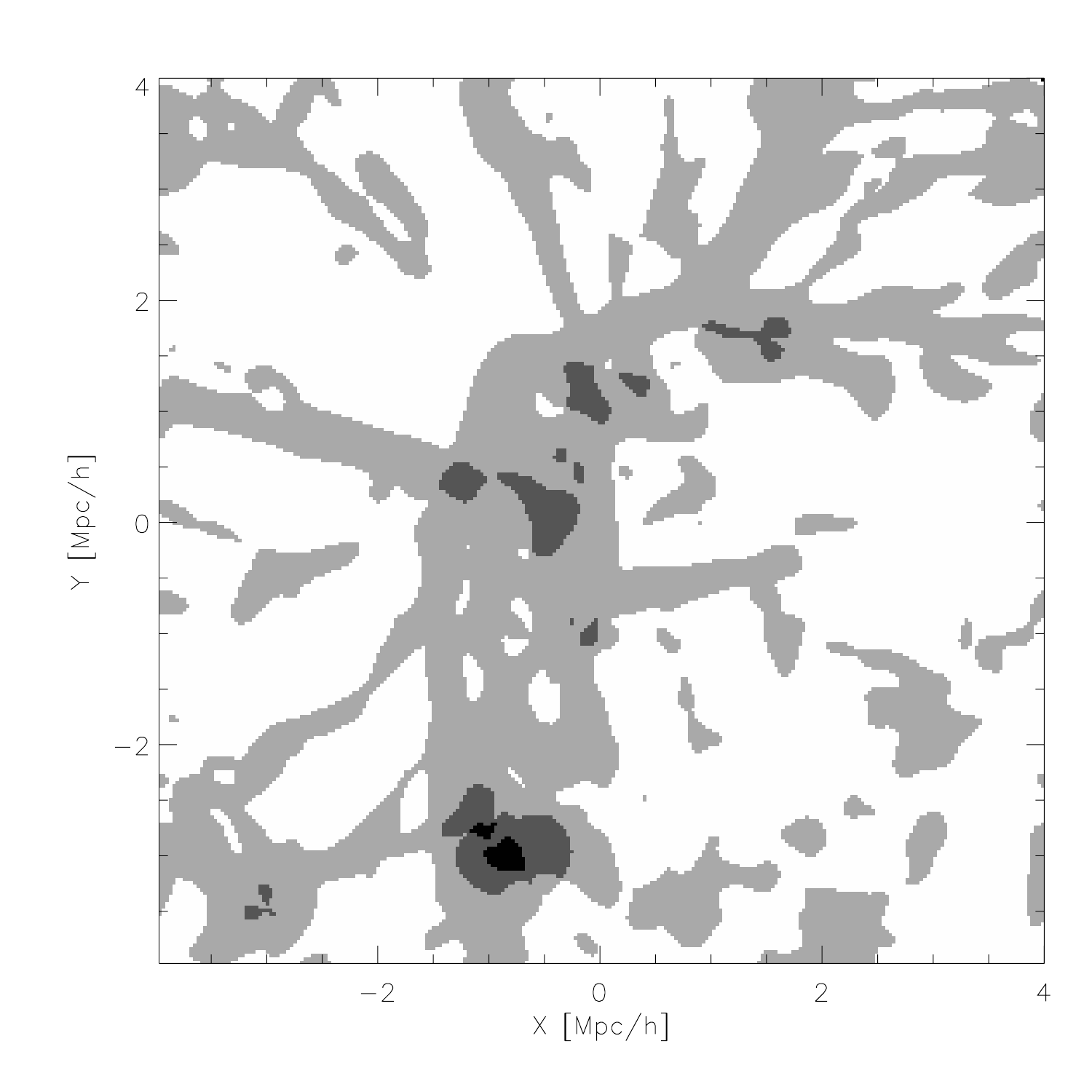}
\includegraphics[width=5.8cm]{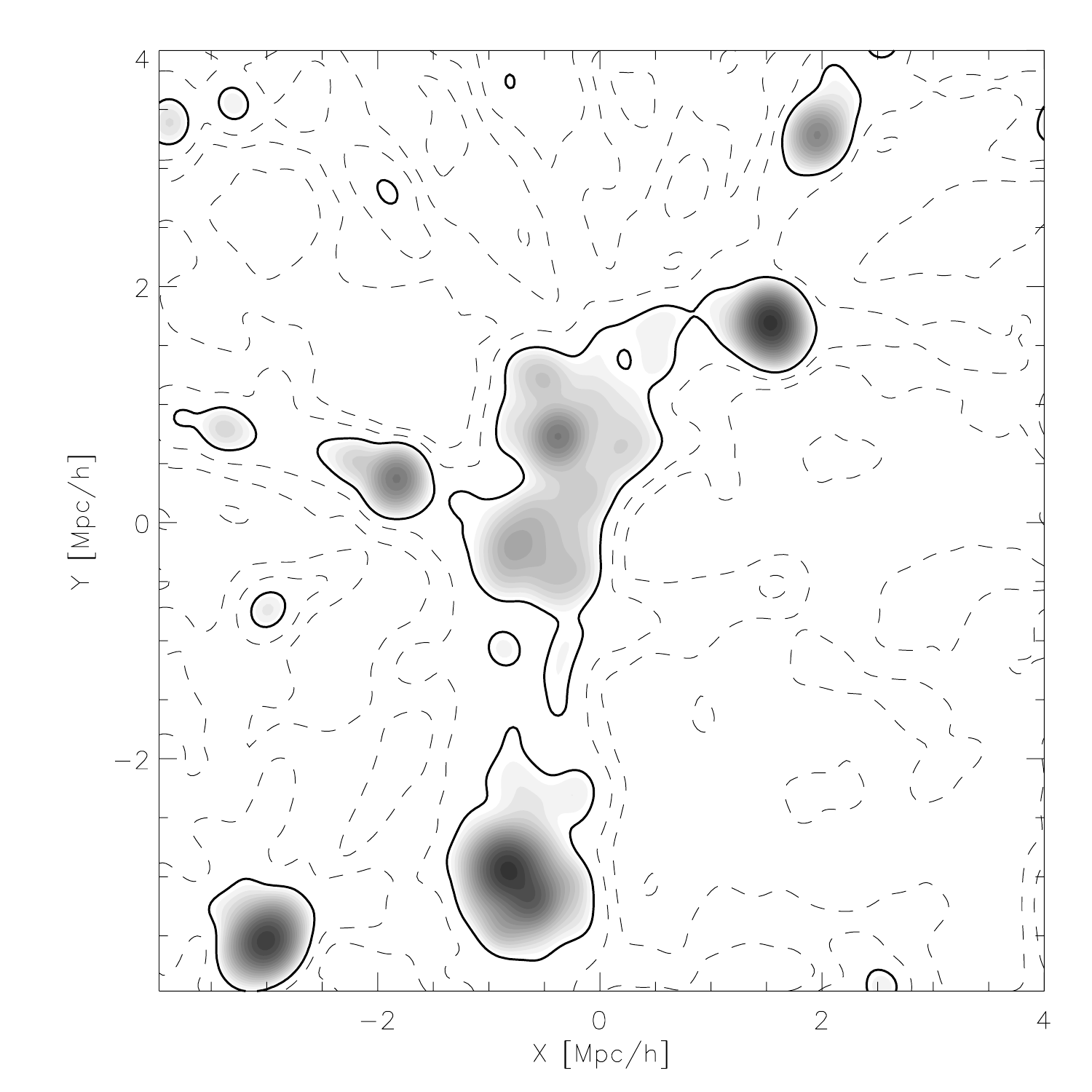} 
\includegraphics[width=5.8cm]{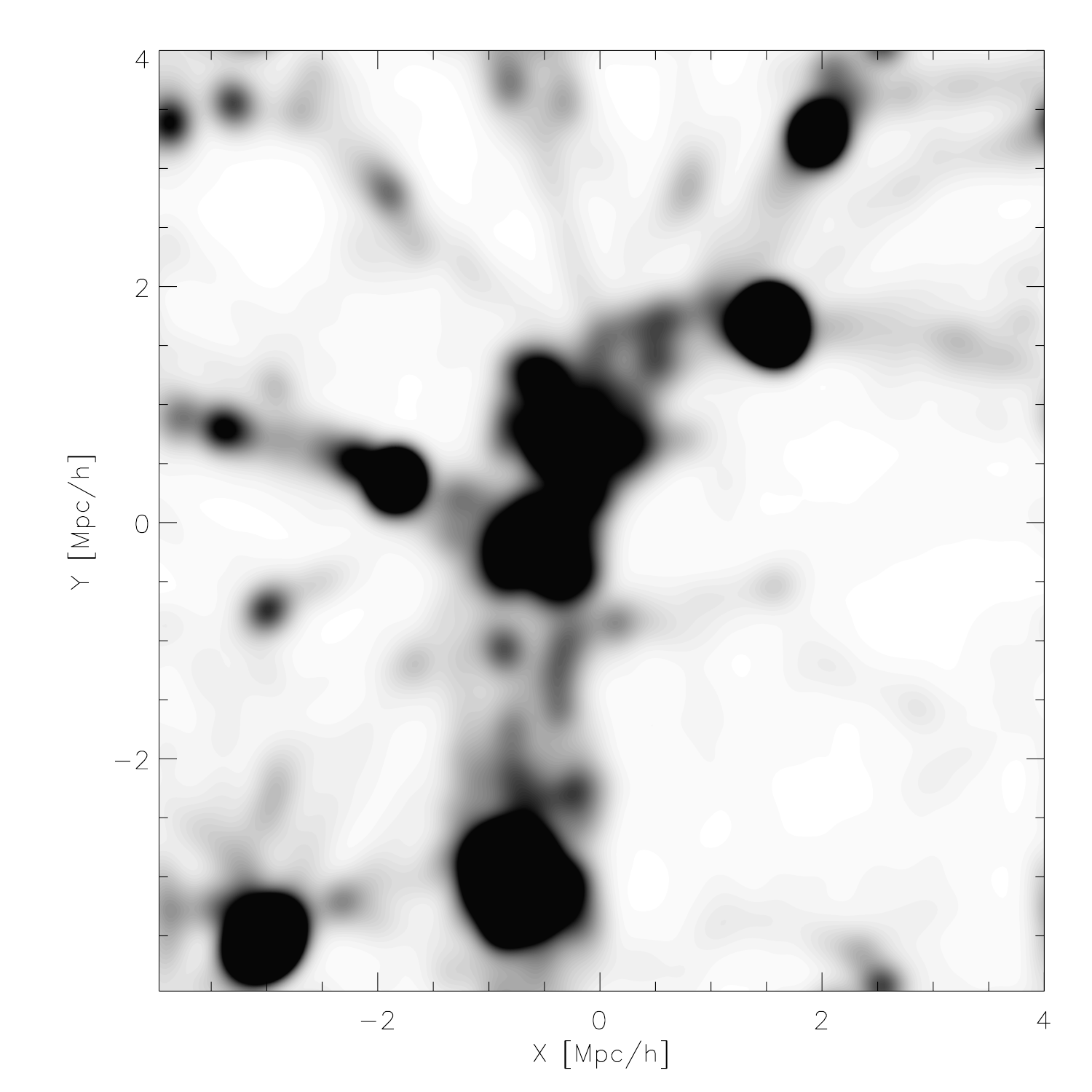}
\includegraphics[width=5.8cm]{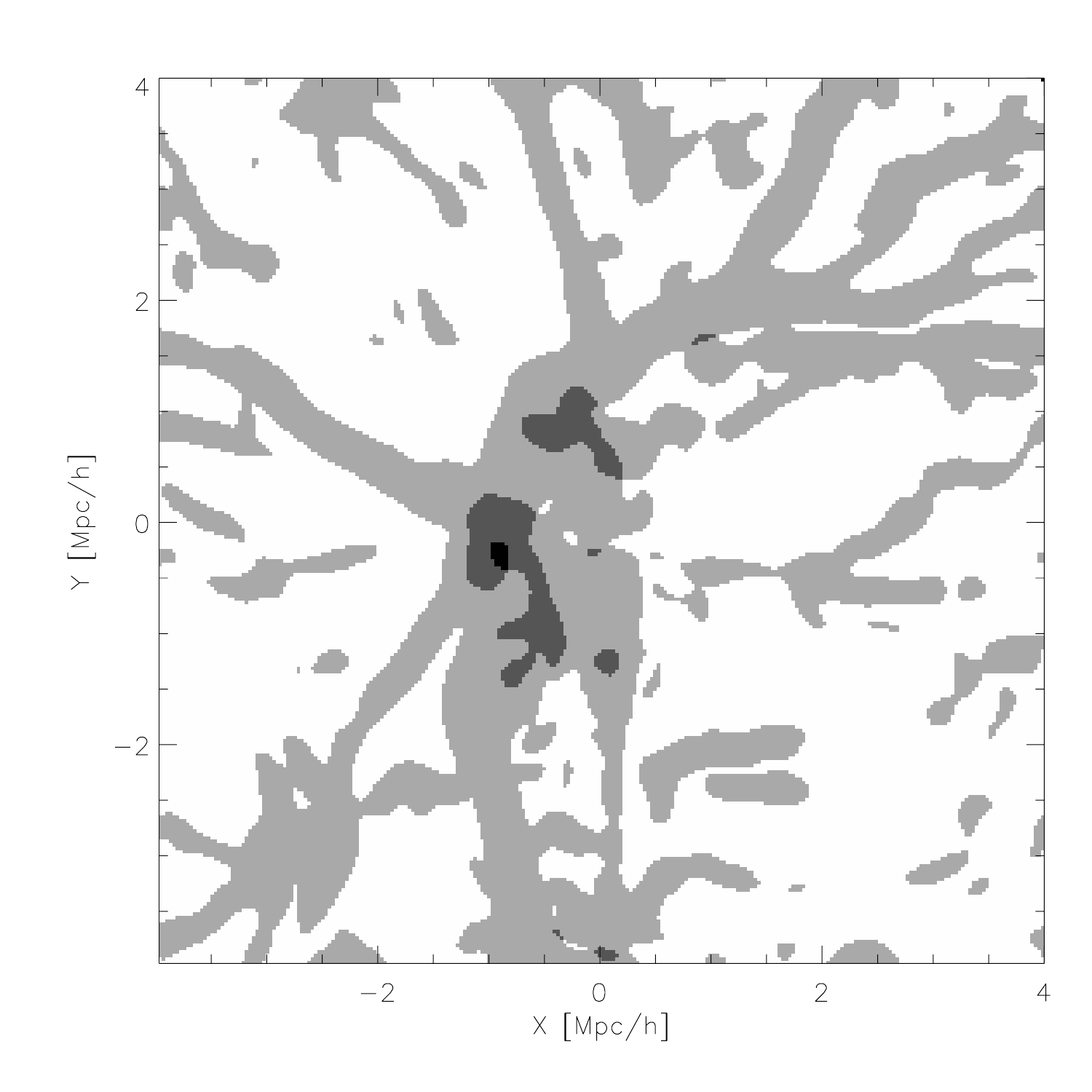}
\includegraphics[width=5.8cm]{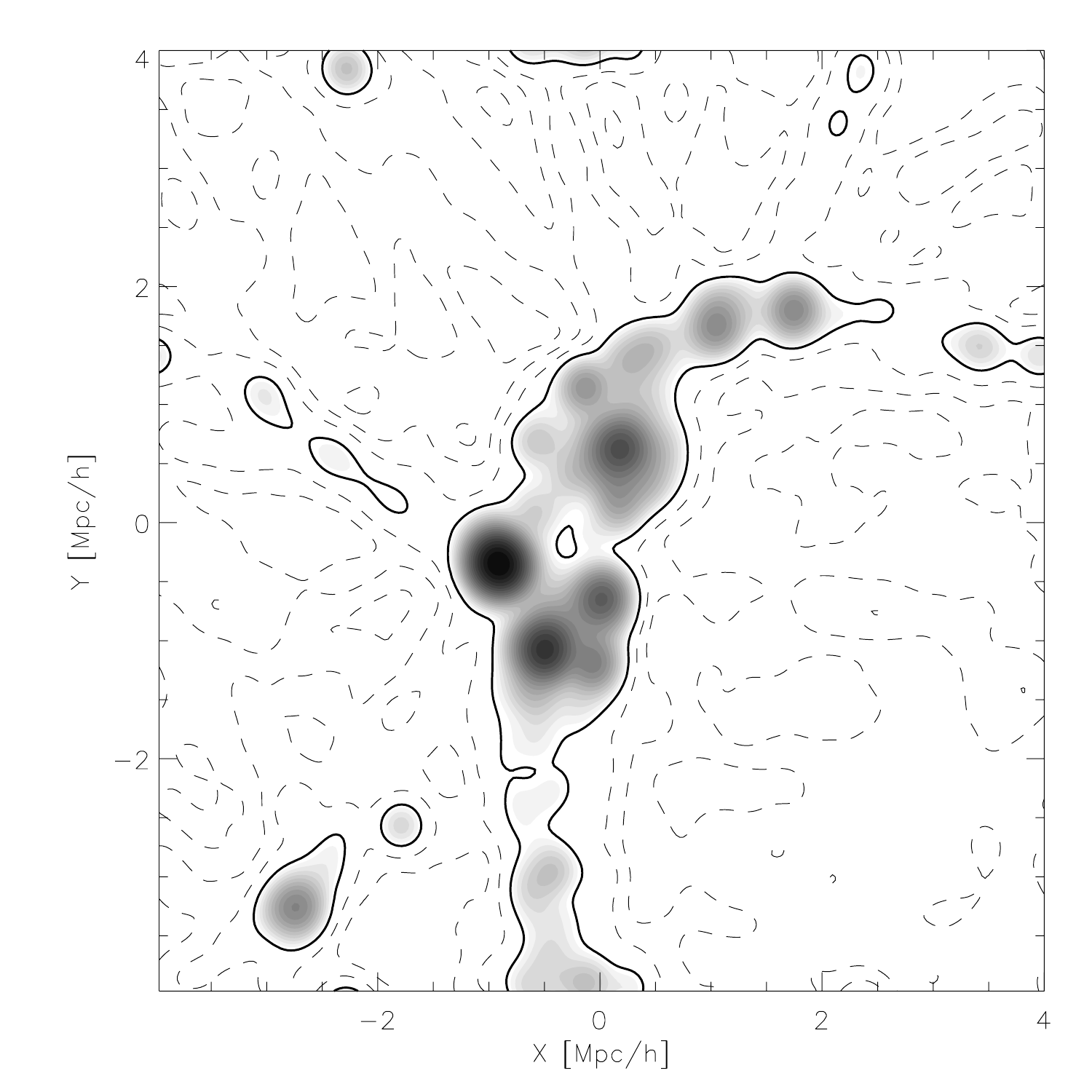} 
\includegraphics[width=5.8cm]{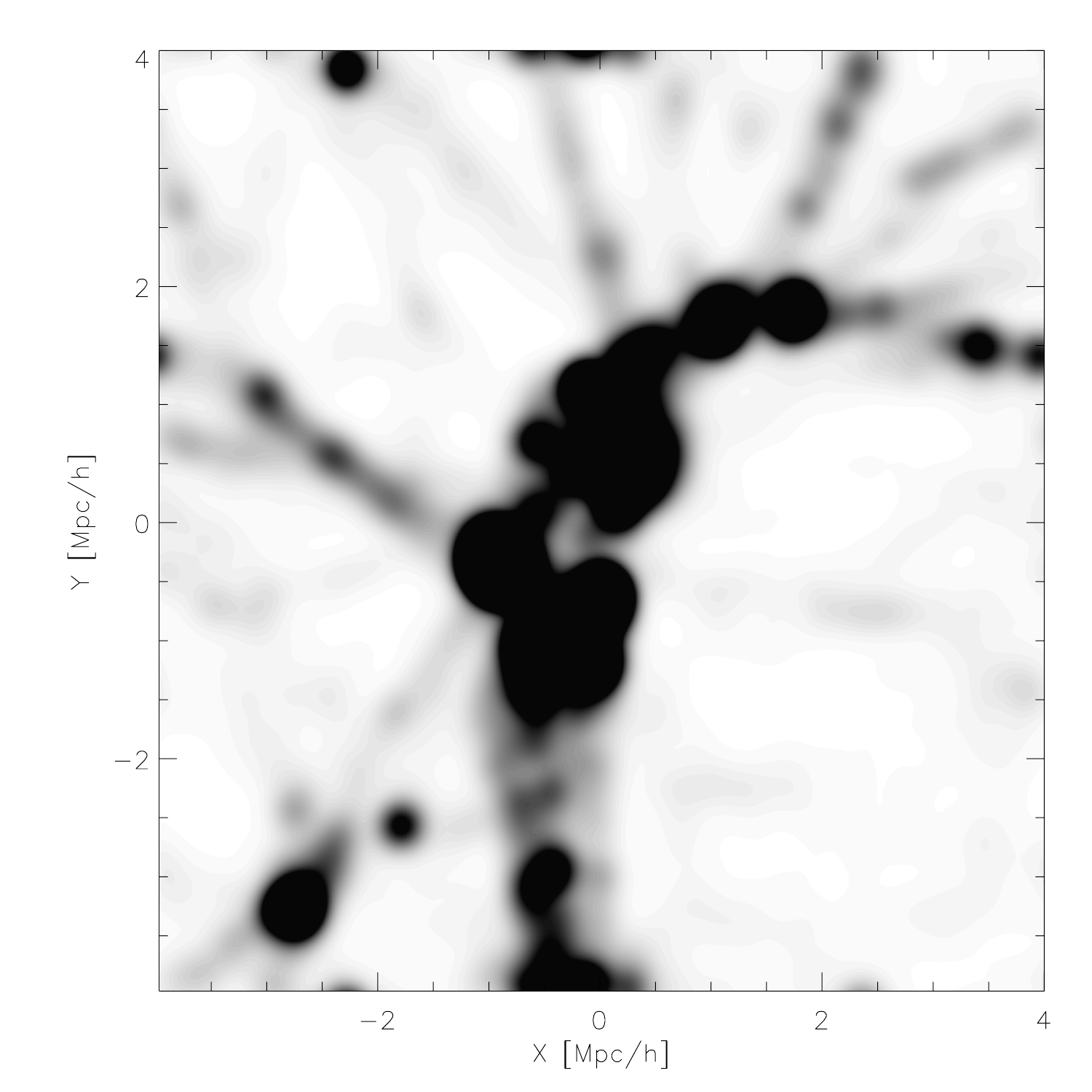}
\includegraphics[width=5.8cm]{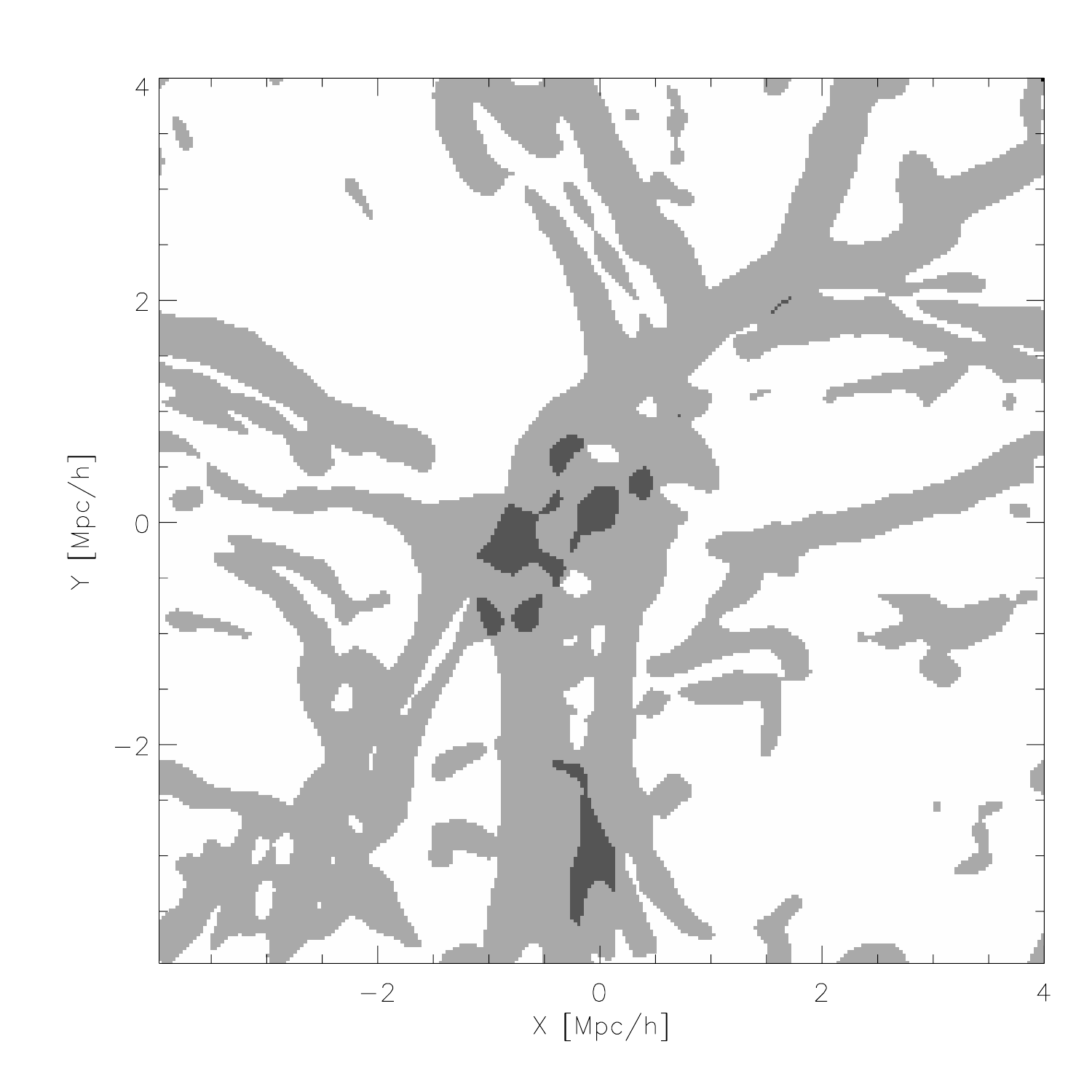}
\includegraphics[width=5.8cm]{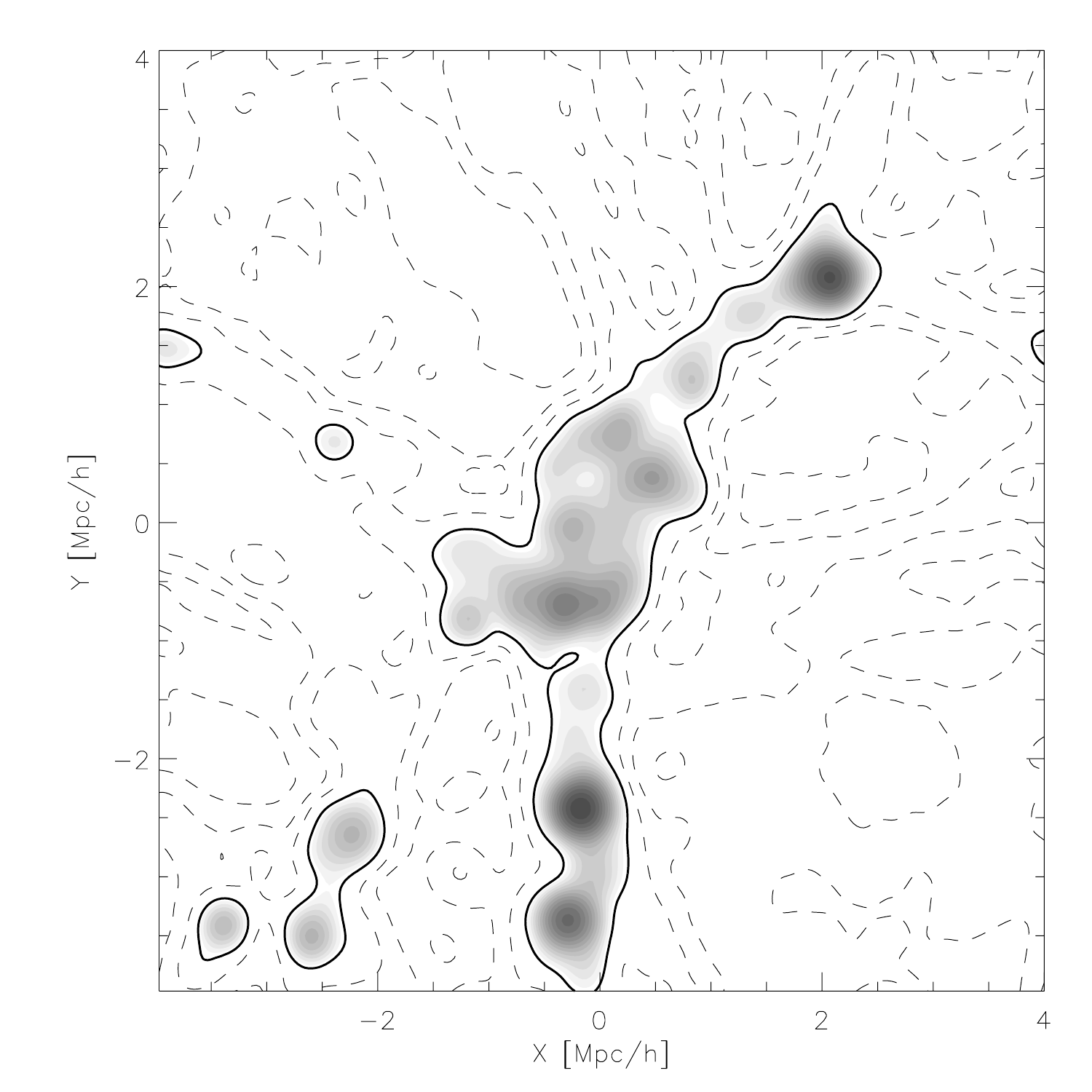} 
\includegraphics[width=5.8cm]{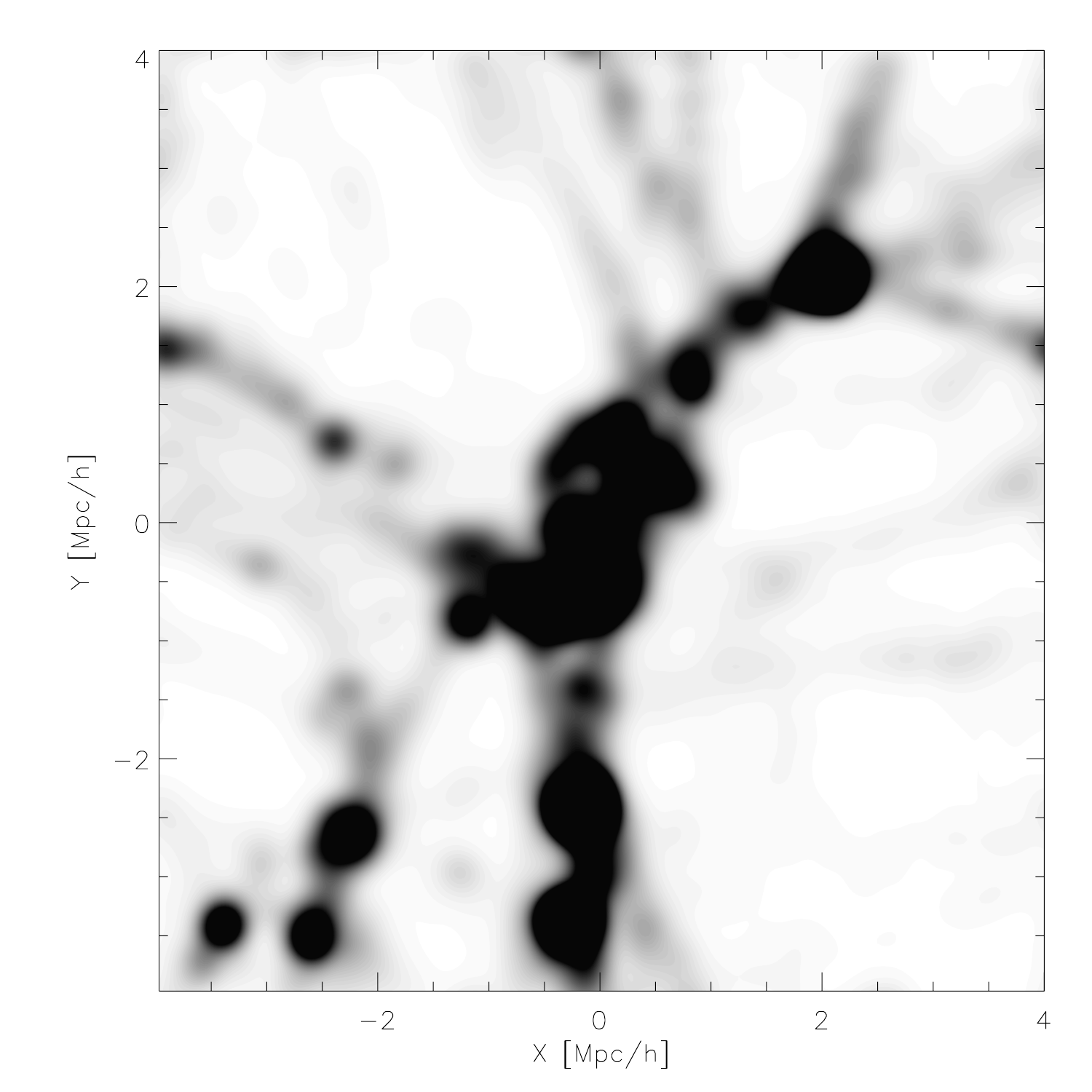}
\caption{Zoom on the center of the computational box  showing the density fields in linear and log scale and the V-web at a Gaussian smoothing of $R_s=0.125\hmpc$. Each row contains the V-web (left), log scale density (middle) and linear density field (right) maps. The upper, middle and bottom rows correspond to the $SGZ=3, 2$ and $1\hmpc$ planes, respectively.
(The color coding and threshold values of Figure \ref{fig:box64-web}  are followed here.)
}  
\label{fig:box8_web_1} 
\end{center} 
\end{figure*}

\begin{figure*}  
\begin{center} 
\includegraphics[width=5.8cm]{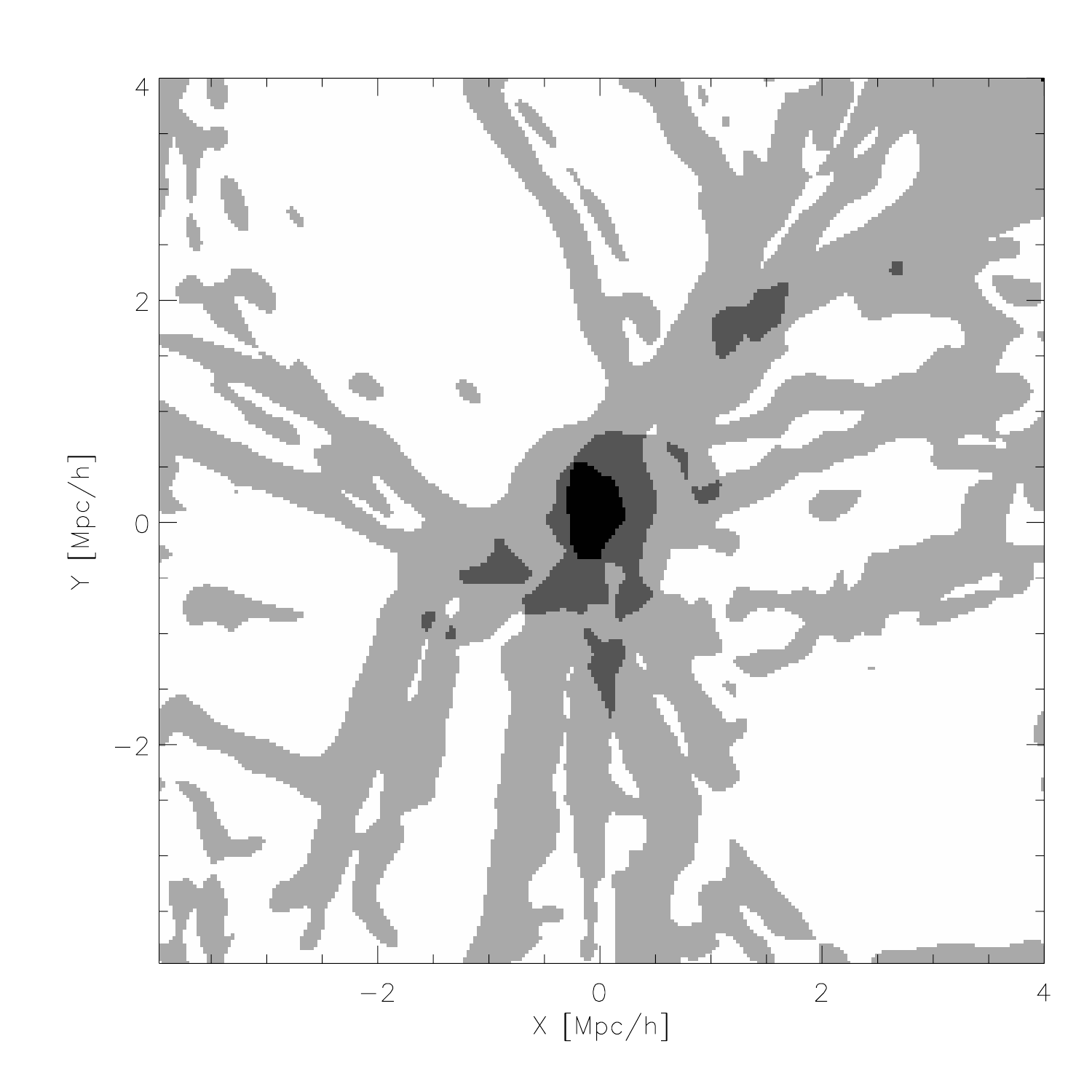}
\includegraphics[width=5.8cm]{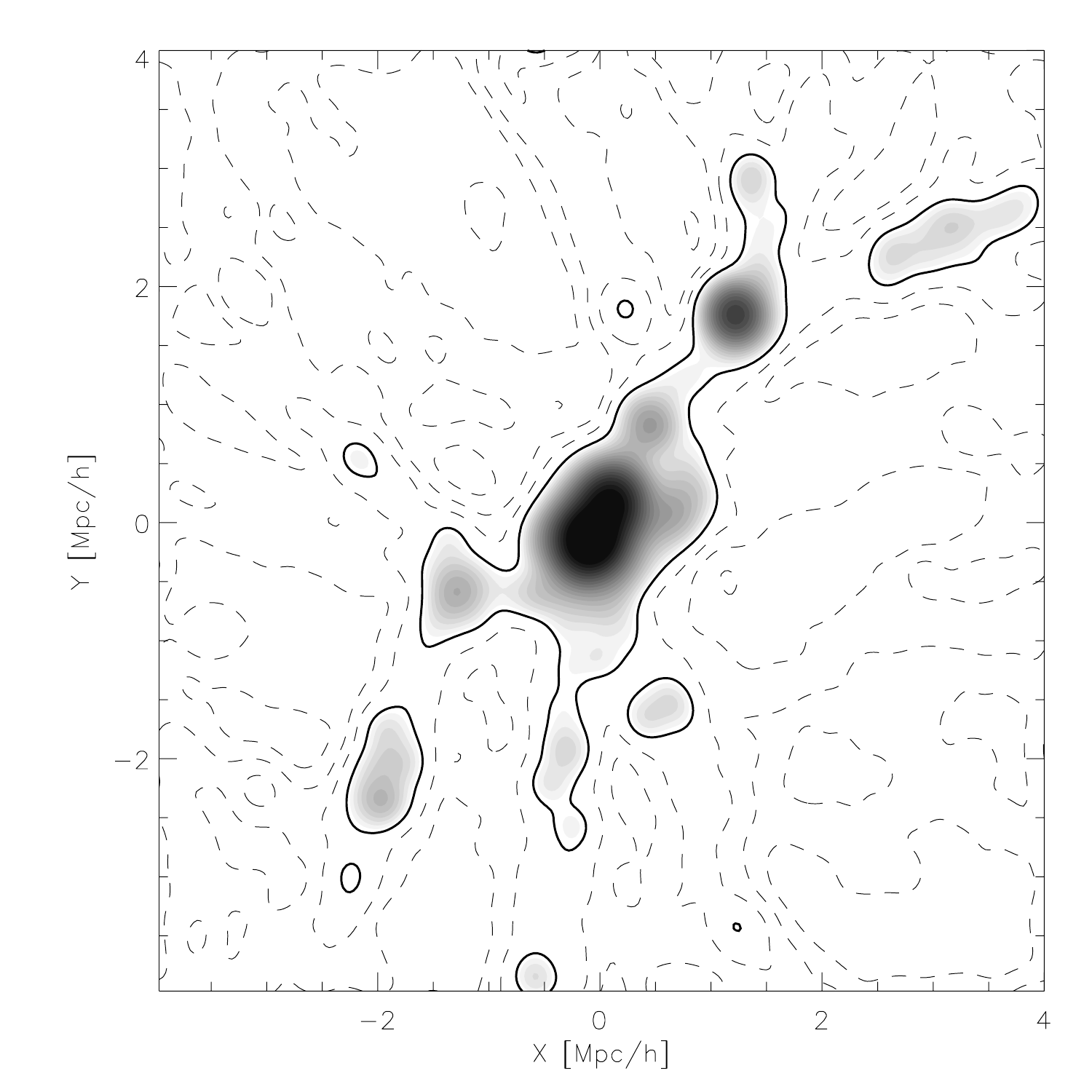} 
\includegraphics[width=5.8cm]{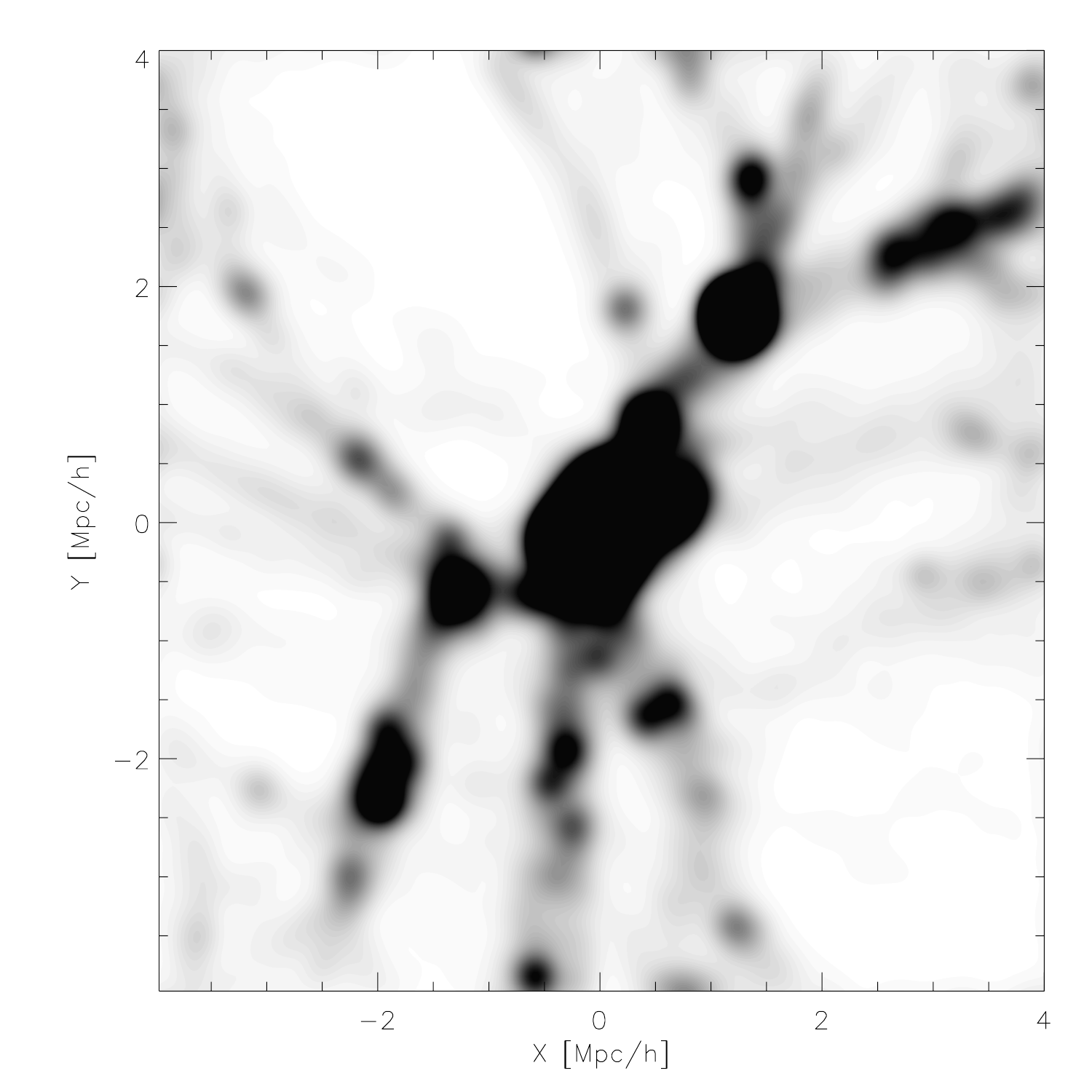}
\includegraphics[width=5.8cm]{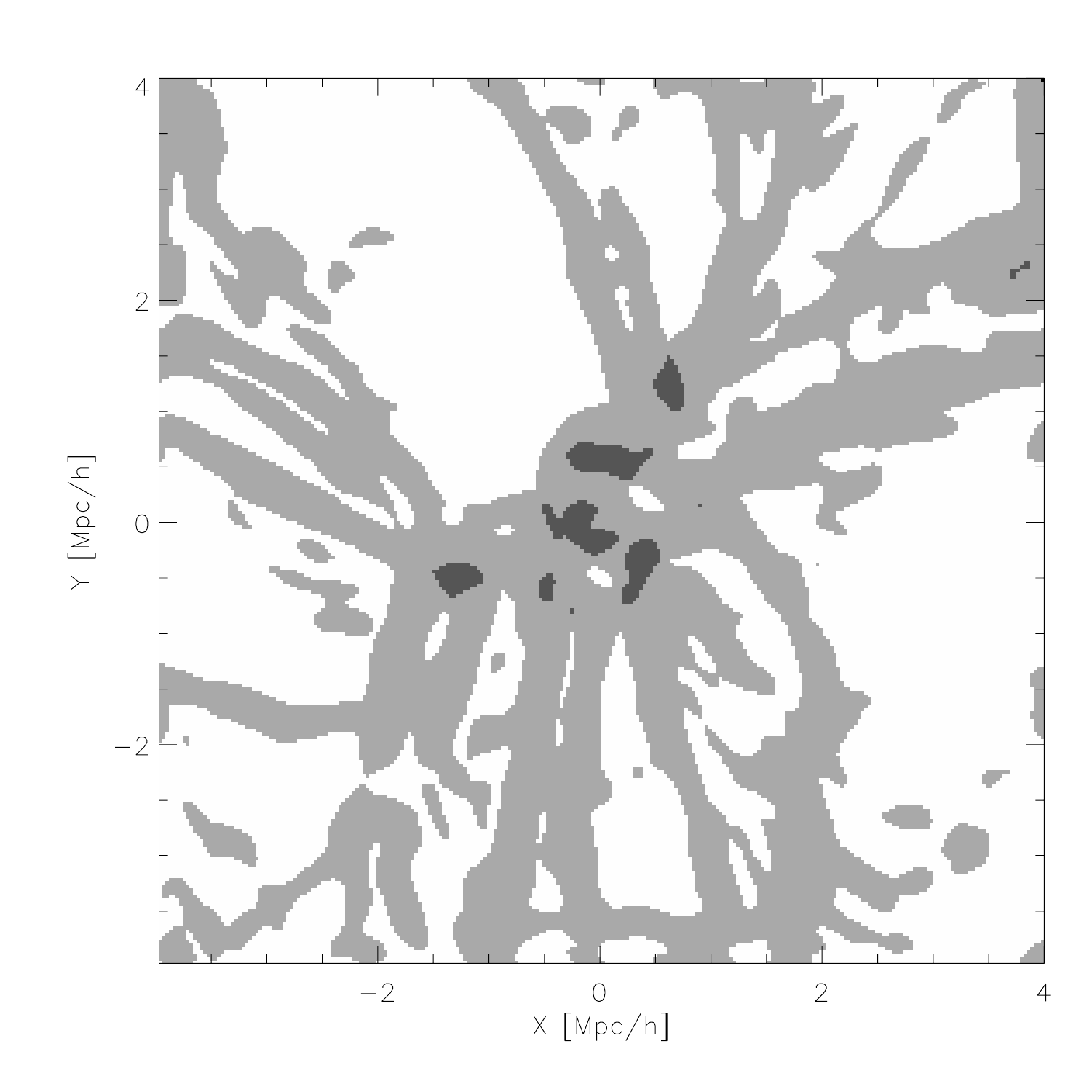}
\includegraphics[width=5.8cm]{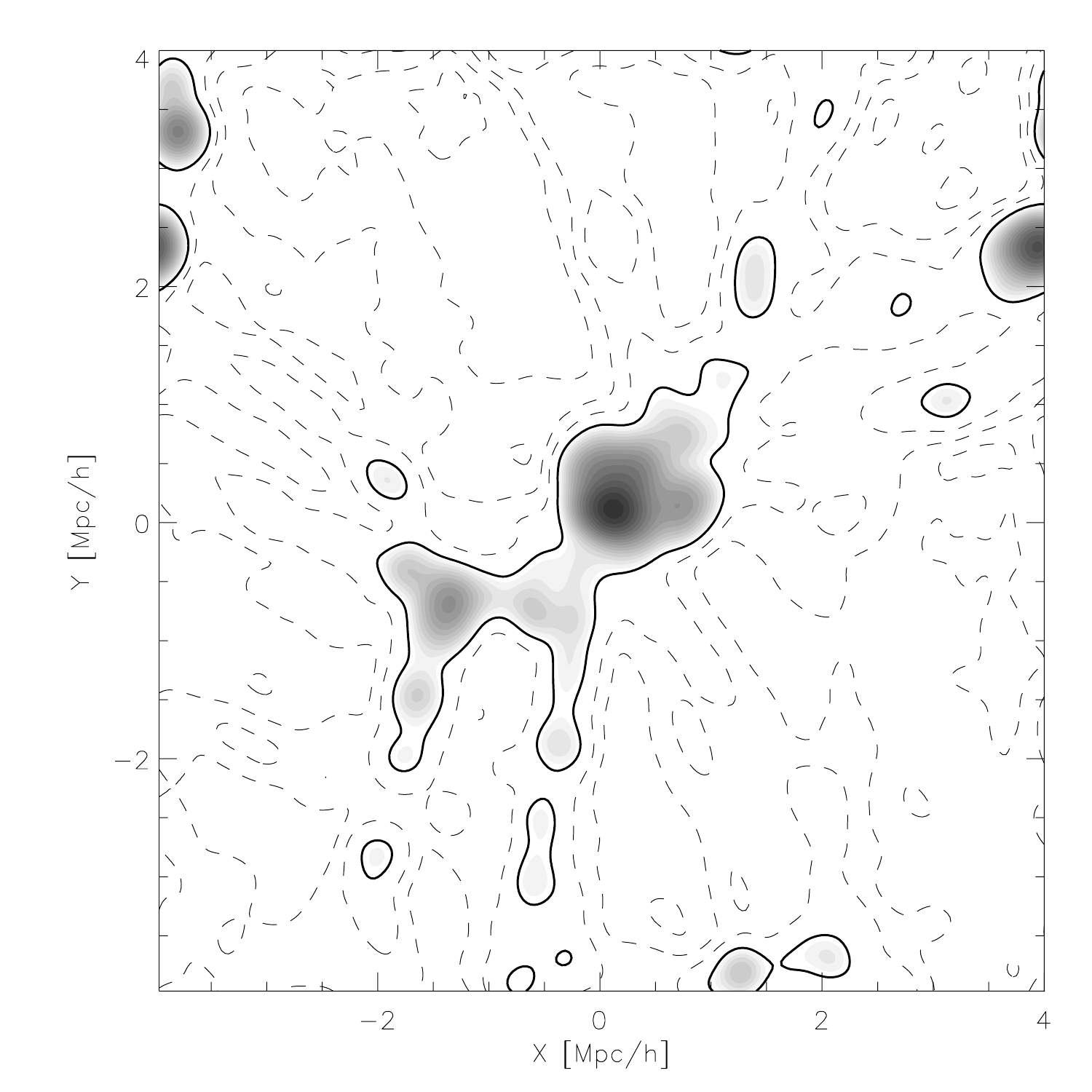} 
\includegraphics[width=5.8cm]{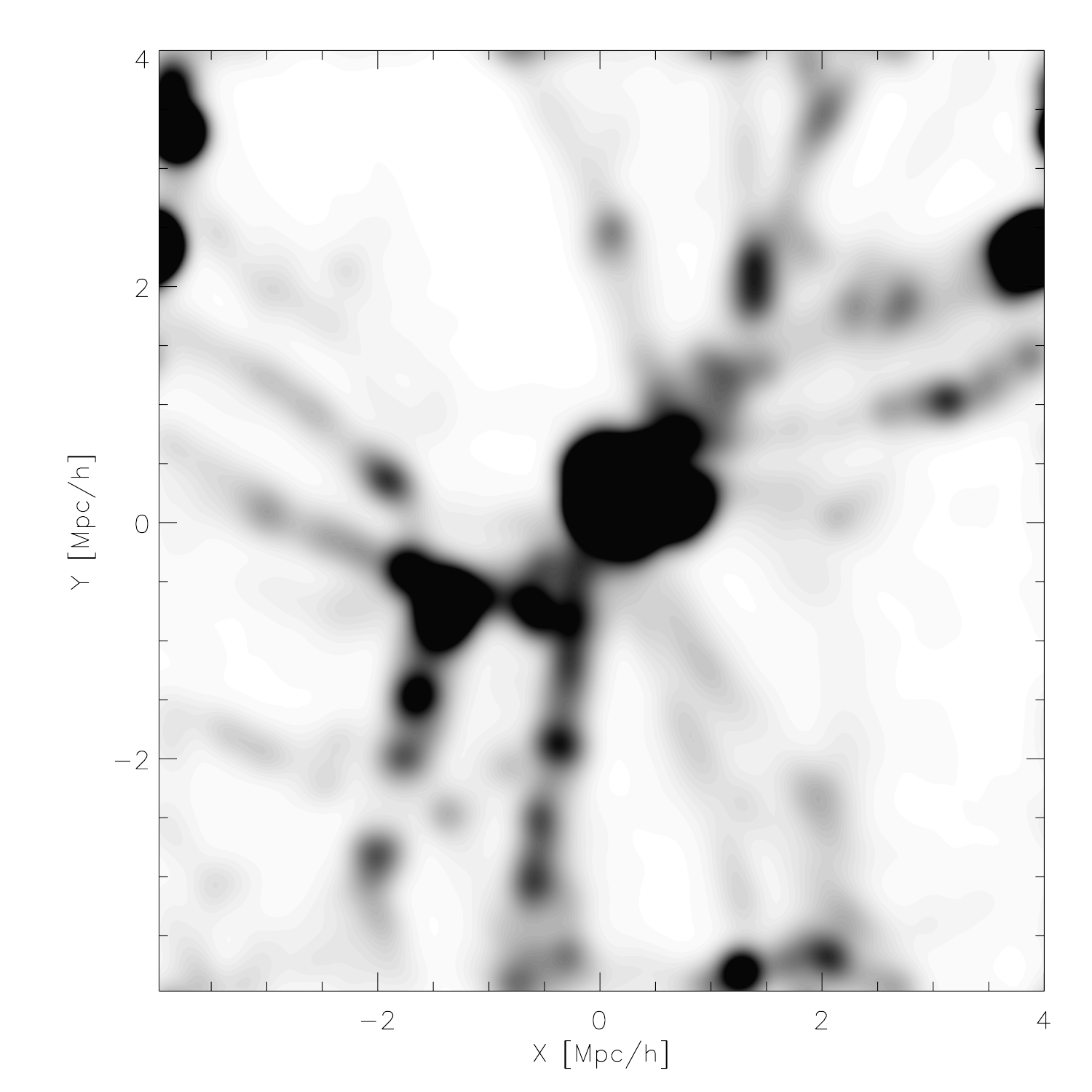}
\includegraphics[width=5.8cm]{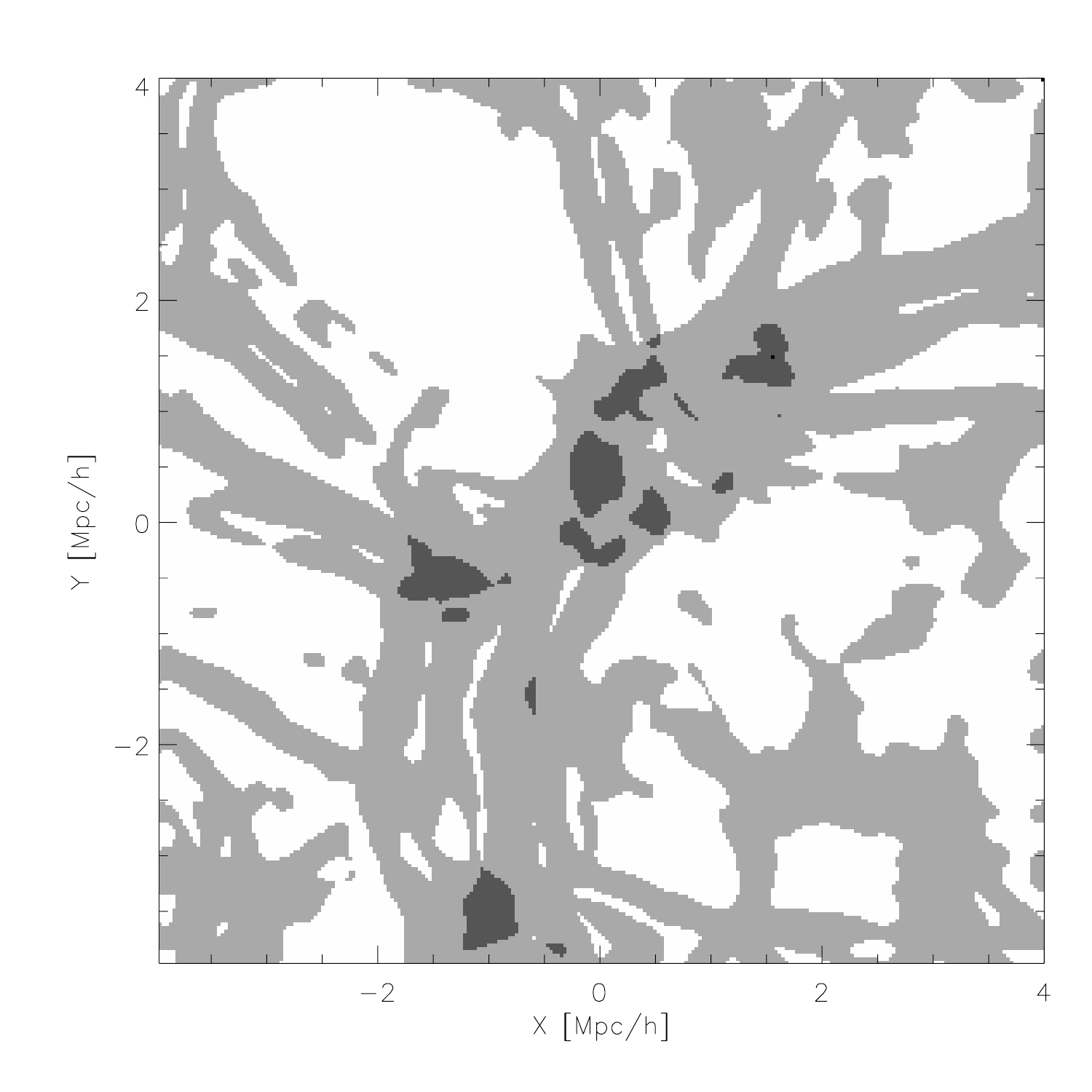}
\includegraphics[width=5.8cm]{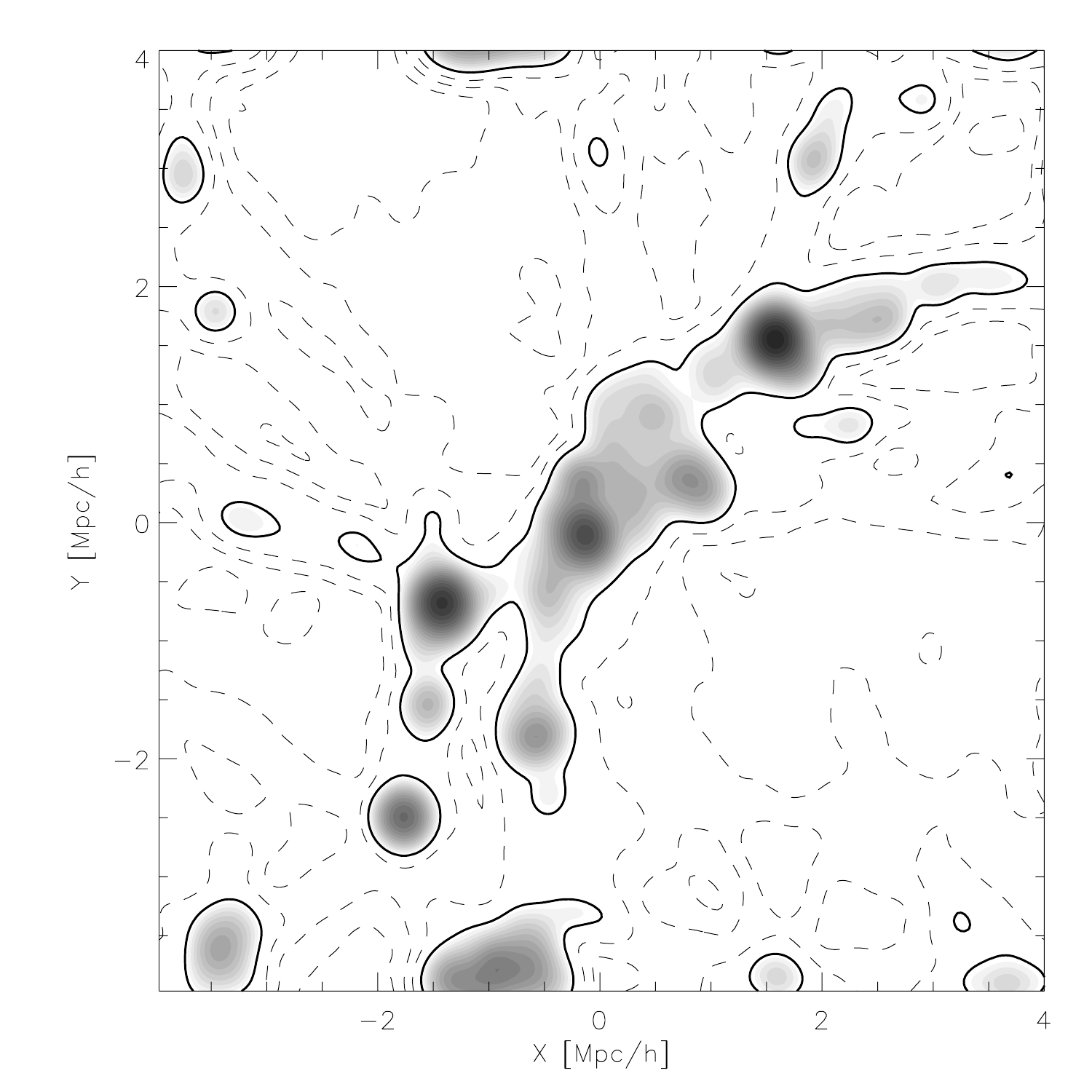} 
\includegraphics[width=5.8cm]{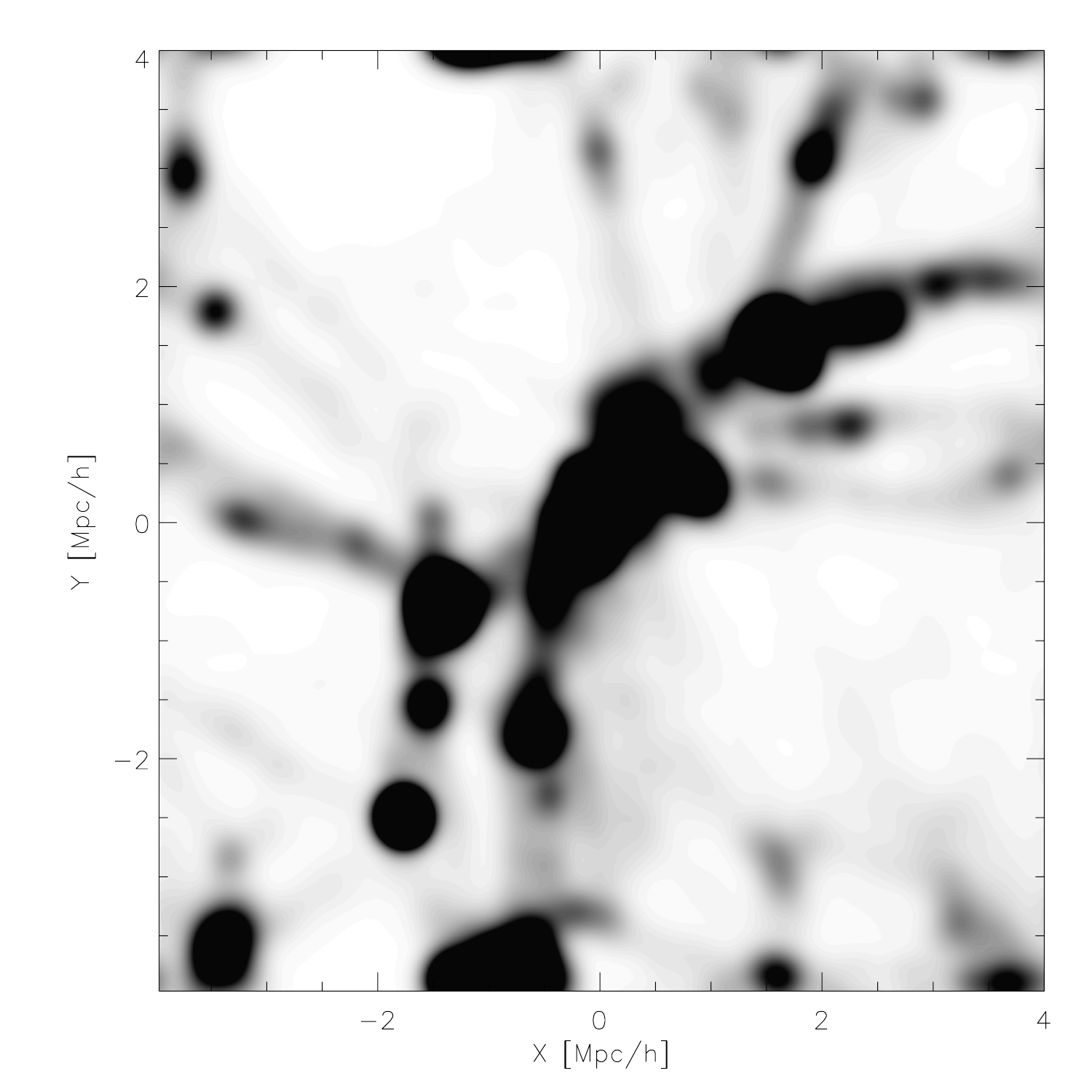}
\includegraphics[width=5.8cm]{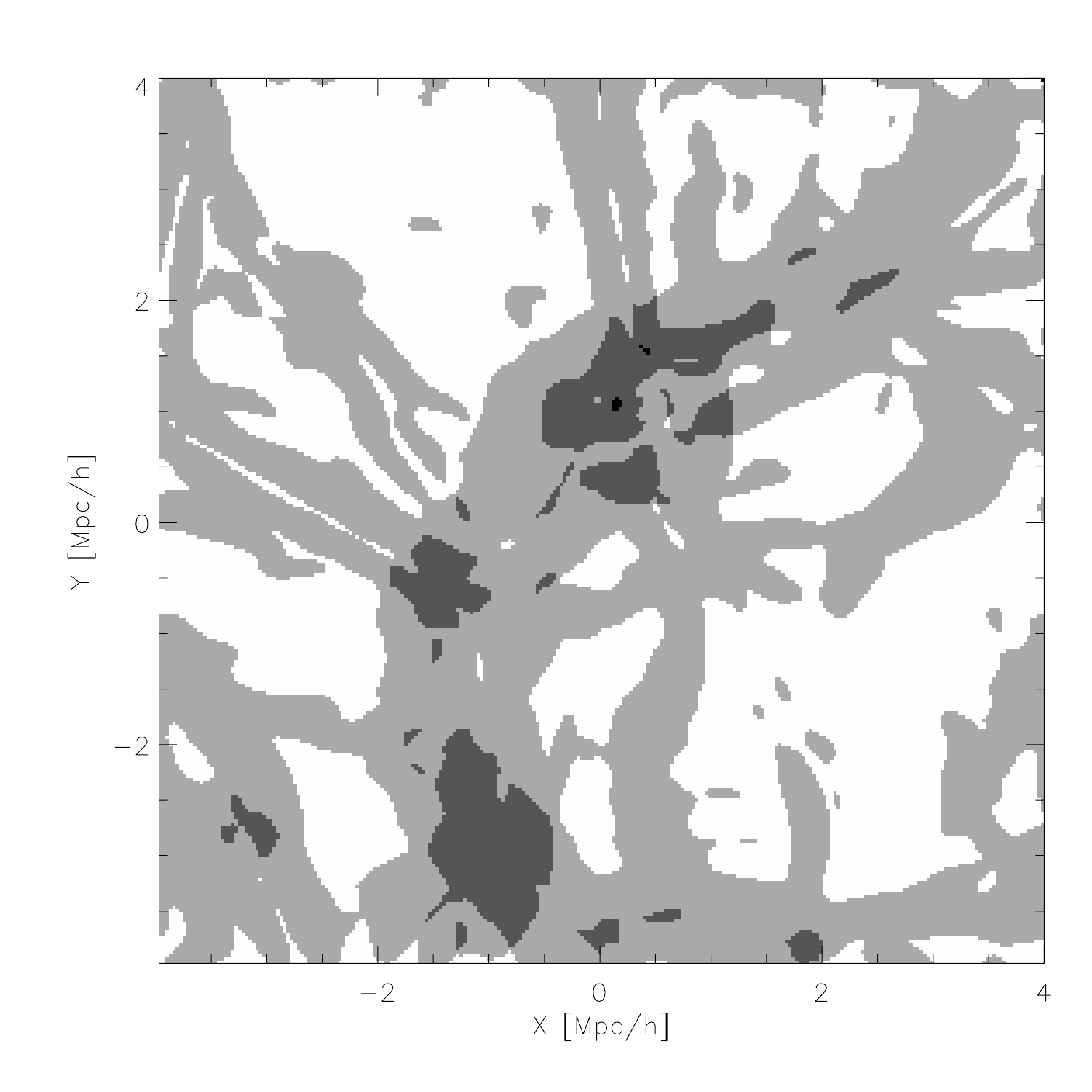}
\includegraphics[width=5.8cm]{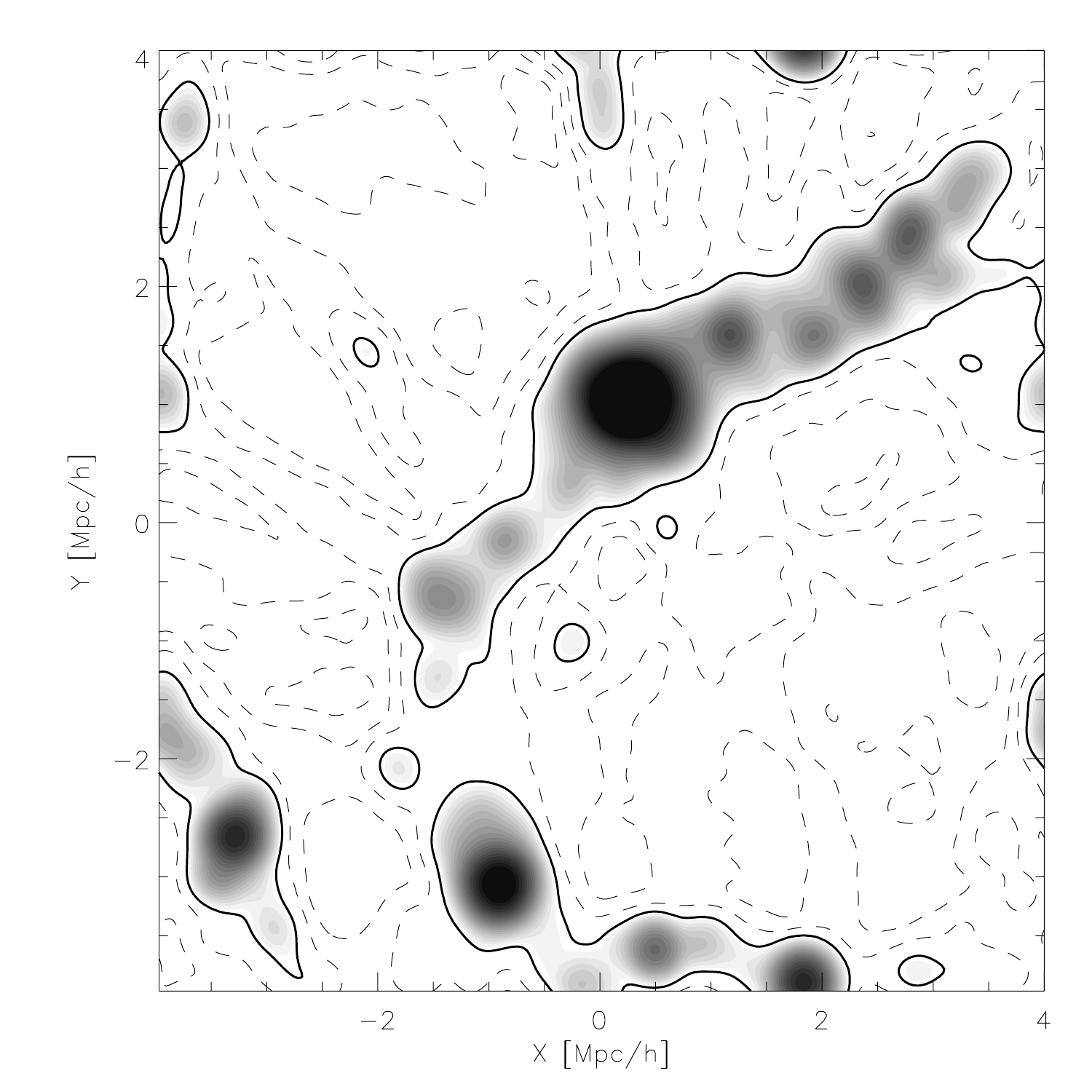} 
\includegraphics[width=5.8cm]{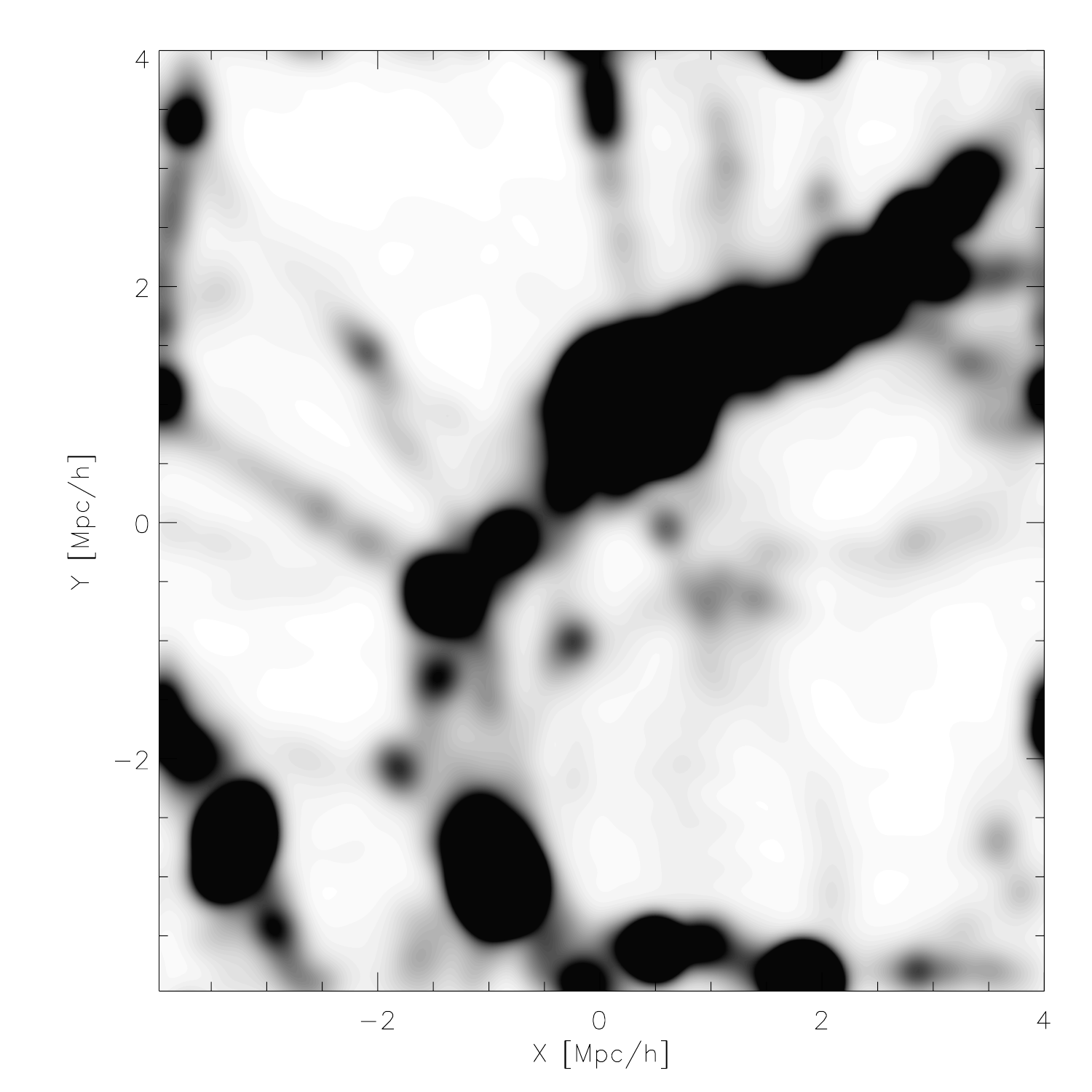}
\caption{Same as Figure \ref{fig:box8_web_1} , for $SGZ=0, -1, -2$ and $-3\hmpc$.
}  
\label{fig:box8_web_2} 
\end{center} 
\end{figure*}

Figures \ref {fig:box8_web_1} and   \ref {fig:box8_web_2} present  a zoom on the inner $(8\hmpc)^3$ of the simulation, showing the density fields in linear and log scale and the V-web at a Gaussian smoothing of $R_s=0.125\hmpc$. The density and velocity fields have been CIC-ed on a $256^3$ grid spanning the zoom region.
The figures show parallel slices of $Z=-3, -2, -1, 0, 1, 2$ and $3\hmpc$ so as to depict the 3D structure.  The color coding of the density (linear scale) field has been chosen so as to make the structure at the $\Delta({\bf r}) \approx 1$ level to be more  apparent. The log scale density maps provide a clear presentation of  the over-dense regions. The 3D structure that emerges from these parallel plots is of a filament that runs roughly parallel to the Z axis across the slices presented here. The width of the filament depends somewhat on the value of the threshold used to define the web. The threshold also dictates whether the filament appears monolithic or  if it breaks into   smaller parallel filaments. That filament is embedded within a curvy wall, i.e. sheet, of $\Delta({\bf r}) \approx 1$ that runs across the slices from $Z=-3$ to $3\hmpc$. The wall leaves its trace on the parallel planes  as a (partial) horseshoe linear structure running from $(X, Z) \approx (-0.5, -4)\hmpc$ through $\approx(0,0)$ towards $\approx(4,3)\hmpc$. That wall is clearly classified by the V-web as a sheet,  it is clearly visible in the (linear scale) density maps.  The wall borders a coherent under-dense region on the lower-right quarter of all the parallel planes. Now this under-dense region, that runs parallel to the Z axis, is not monolithic. It is further split by density ridges, forming some more smaller scale sheets that  cut through the under-dense big blob creating smaller voids. These ridges are under-dense themselves, yet they are denser than the voids they surround. Some other coherent voids, sheets and filaments   that run perpendicular to the $Z=const.$ planes are apparent as well.

\begin{figure*}  
\begin{center} 
\includegraphics[width=7.5cm]{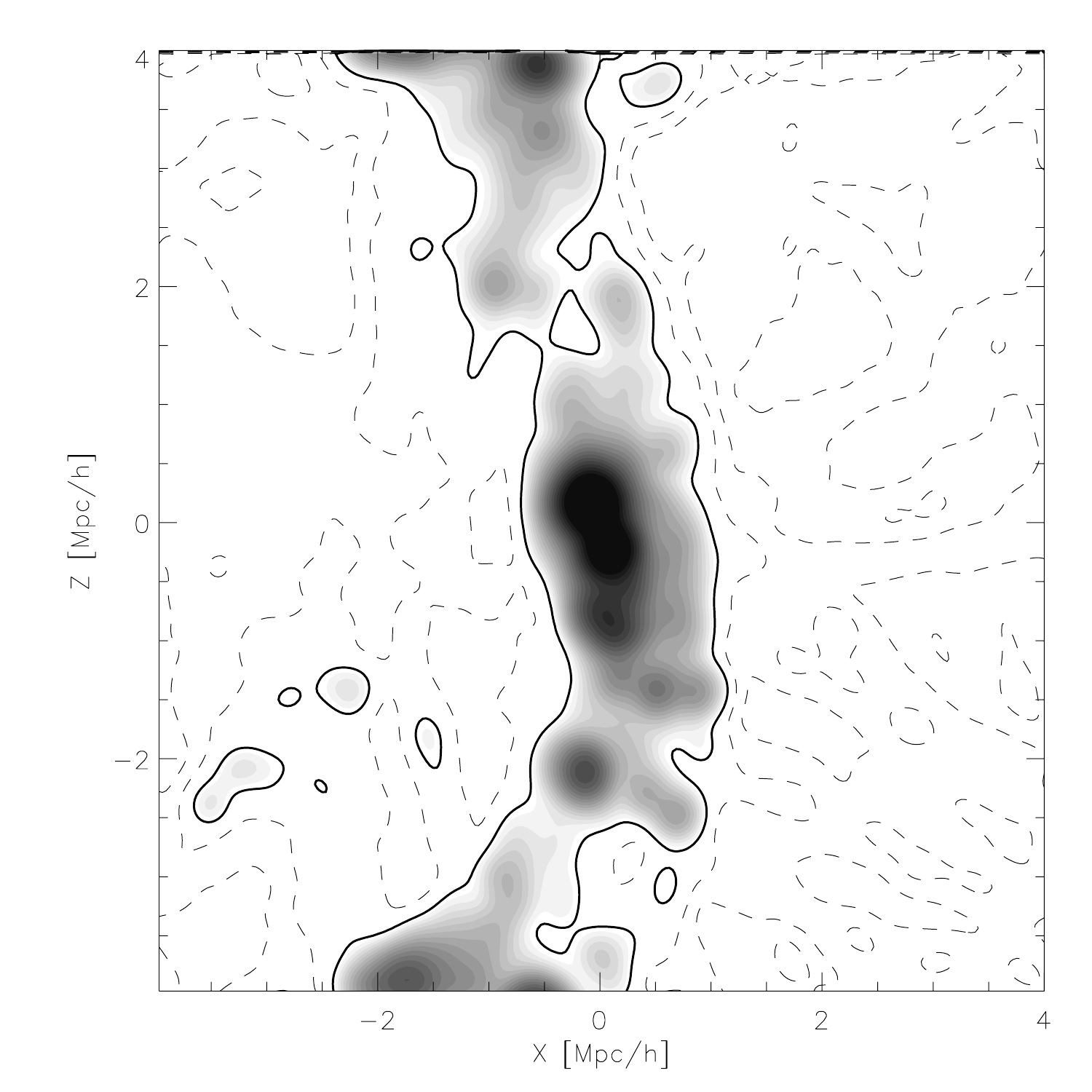}
\includegraphics[width=7.5cm]{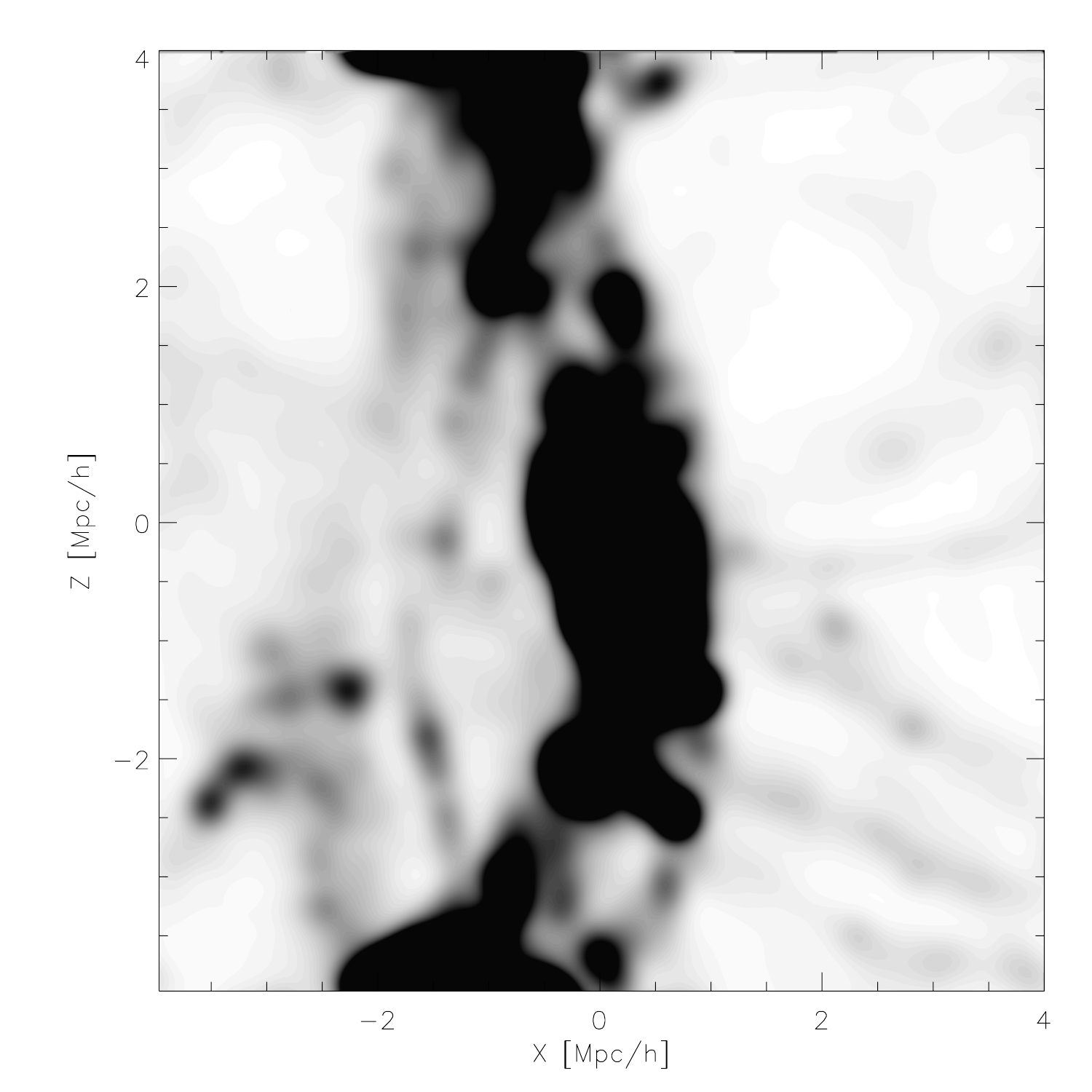} 
\includegraphics[width=7.5cm]{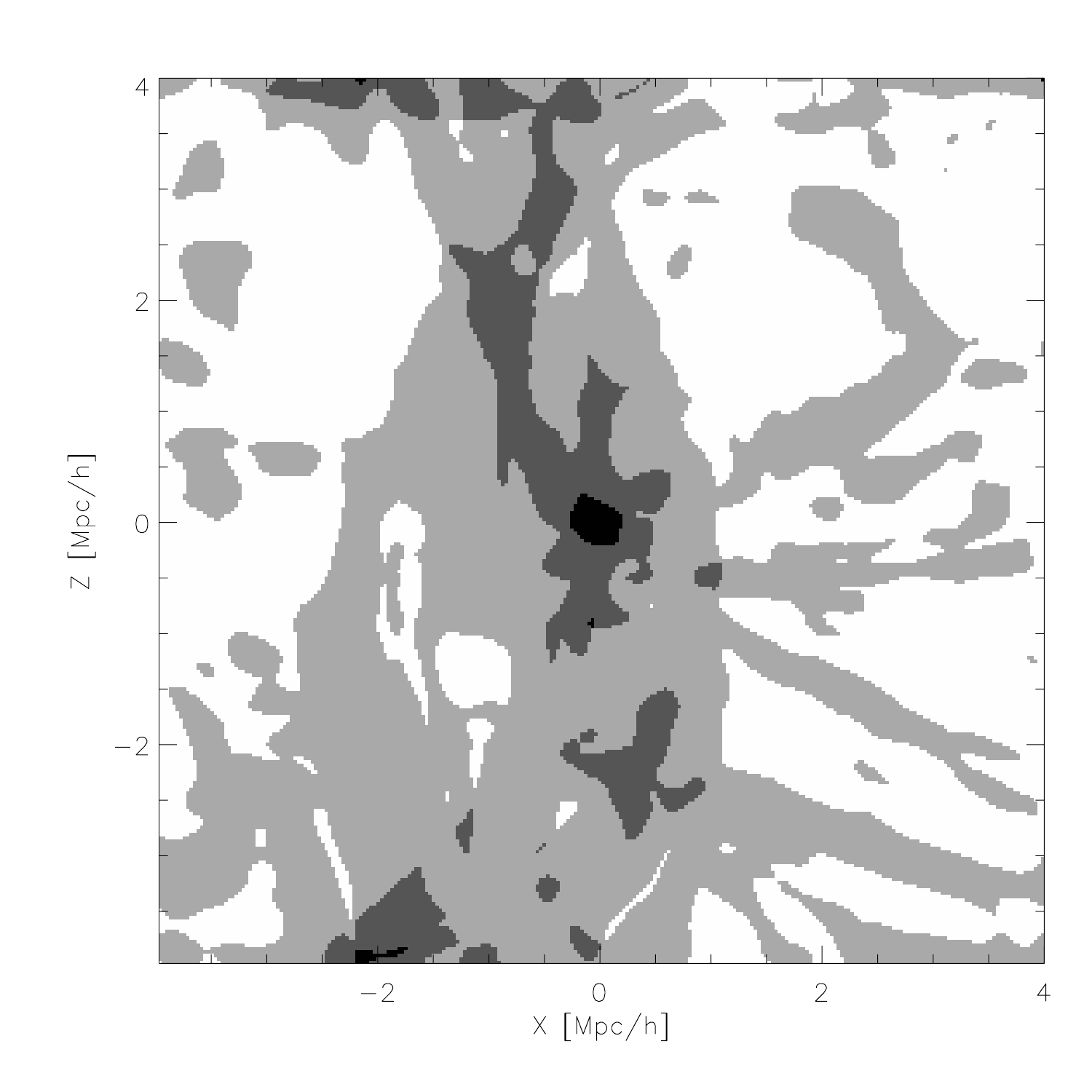}
\includegraphics[width=7.5cm]{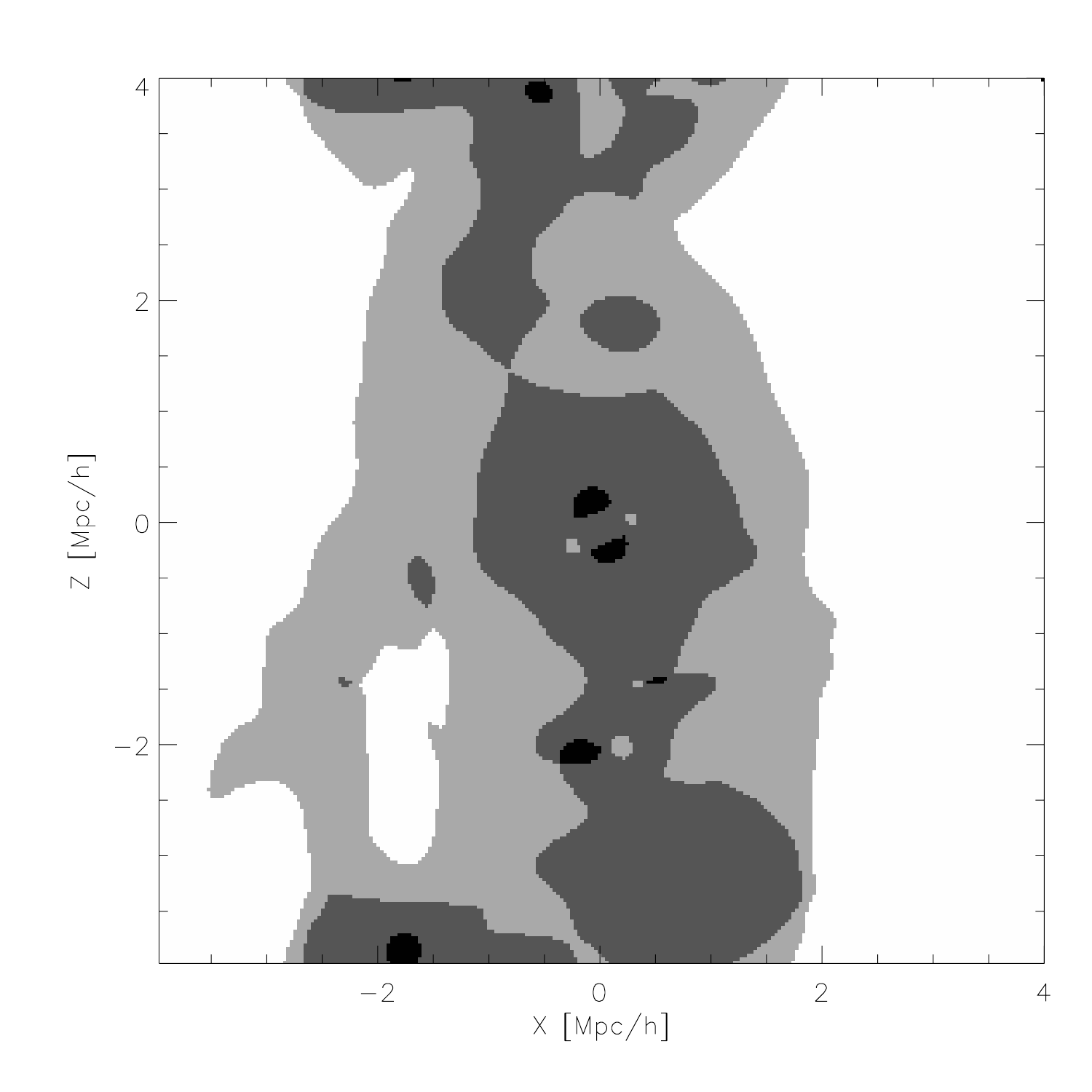}  
\caption{ A comparison of the V-web and T-web of the XZ-plane of the zoom box (shown in Figures  \ref {fig:box8_web_1} and  \ref {fig:box8_web_2}, using the same resolution and threshold values of Figure \ref {fig:box64-web}: 
a.The density field presented by $\log\Delta$ (upper-left panel).
b. The linear density field   (upper-right panel). 
c. The velocity based cosmic web    (bottom -left panel).
d. The (gravitational) tidal   cosmic web (bottom-right panel).
The color and contour coding of Figure \ref {fig:box64-web} is followed here.
}
\label{fig:box8_web} 
\end{center} 
\end{figure*}

 The two mid-panels of Figure \ref {fig:box64-web}, which depict the V- and T-web, clearly show that the improved spatial resolution of the V-web compared with the T-web.  The comparison is further extended in Figure \ref{fig:box8_web}.
Again, lacking a quantitative measure  of the quality of the construction of the cosmic web we resort here to a visual comparison of the constructed cosmic web with the underlying density field. The plots show the XZ plane of the zoom box presented in the series of the Z cuts of Figures \ref{fig:box8_web_1} and 
 \ref{fig:box8_web_2}.  The sheet that almost coincides with the YZ principle plane is clearly manifested in the orthogonal XZ cut. The plots also show the filament that runs very close to the Z axis. The fine details, on sub-Megaparsec scale,  of the density field are much better reproduced by the V-web than by the T-web.

\section{Some statistics and  a multi-scale approach  }
\label{sec:multi}

A close inspection of the the density and V-web maps (Figures  \ref{fig:box64-web} -  \ref {fig:box8_web}) shows a very clear correlation of the V-web classification with the local density. There is clear correspondence between the web type with density, with voids, sheets, filaments and knots progressing from the most under-dense to the densest environments, respectively. 
Yet, there is no one-to-one correspondence between the web type and density, as is manifested by Figure 5, whose left panel shows the probability  distribution of grid cells   for the different web elements as a function of the fractional density, $\Delta$.  The plot shows that virtually at all density levels cells of any web type can be found. In particular, some  very dense cells with $\Delta\approx 10^3$ are tagged as voids.  A close inspection reveals that all such points are located at the very inner regions of massive halos, where the velocity field reflects the virial motions within virialized halos. This is a clear artifact of the over-resolution of the CIC grid which penetrates too deeply into massive halos. Only a very few cells are affected by this over-resolution and the global statistics is virtually unaffected. A similar effect is found also for some of the cells with density exceeding  the virial mean density ($\Delta\approx 340$ for the cosmological parameters assumed here), which are classified as sheets and filaments, and are clearly located within the very inner regions of massive halos. The over-resoluton conjecture is easily verified by constructing a lower resolution V-web and confirming that the pathology disappears with the coarser grid. 

 The following multi-scale procedure effectively removes the pathological misclassified cells. A low resolution V-web is constructed, in addition to the finest desired web. All high resolution cells defined as voids and have an over-desity exceeding unity, $\Delta  \ge   1$, are given the web attributes of the coarse web. Similarly, all high resolutions cells identified as sheets and filaments whose over density exceeds the virial density assume the web classification of the coarse grid. 
 Both the high and law resolution webs are evaluated on the same $256^3$ grid and the resolution is controlled by a Gaussian smoothing.
 The coarse grid is defined here by a Gaussian smoothing of $R_g=1 \hmpc$. The probability distribution  of grid cells  of the corrected web is shown by the right panel of Figure 5.  The corrected V-web is presented by the bottom-left panel of Figure \ref{fig:box64-web}. From here on the V-web is assumed to be corrected by the  multi-scale approach, unless it is otherwise explicitly stated. 

Table 1  presents the volume and mass filling factors of the V-web. The numbers in parentheses correspond to the V-web constructed without the multi-scale correction. The numbers presented here clearly depend on the threshold level and  resolution. Given that the V-web calculated here closely matches the visual appearance implied by the mass distribution, the following remarks are of interest. Voids and sheets occupy $95\%$ of the simulated volume, and only less than $1\%$ is occupied by knots.  The volume and  mass filling factors of the  sheets and filaments are in close agreement with the findings of FR09, for a threshold of unity.  The sheets are more dominant by volume and mass in the V-web compared with the T-web.

\begin{figure*}  
\begin{center} 
\includegraphics[width=7.5cm]{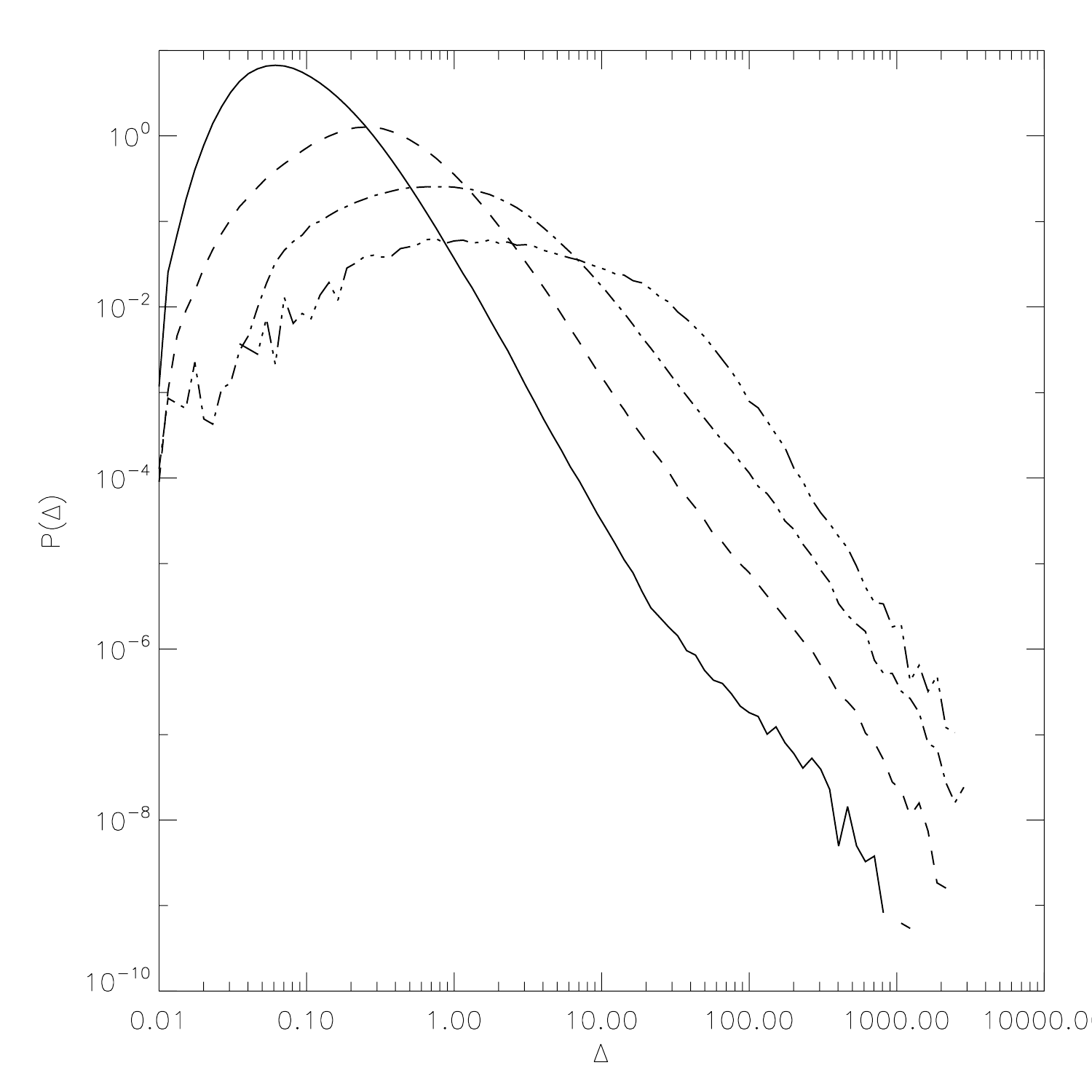}  
\includegraphics[width=7.5cm]{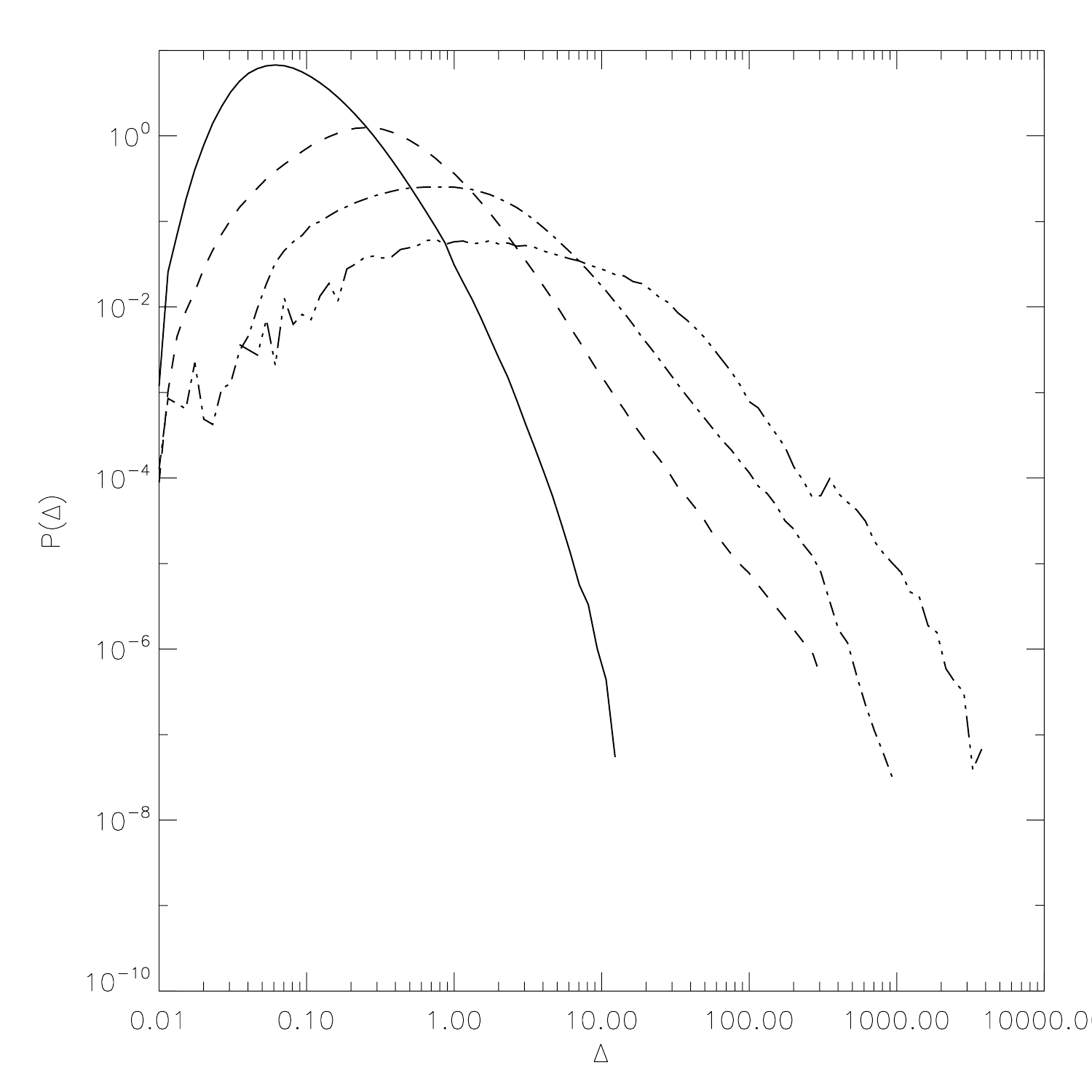} 
\caption{The  probability  distribution of grid cells  as a function of the fractional density,   $P(\Delta)$,   is plotted   for the various V-web elements, voids (full line), sheets (dashed), filaments (dot-dashed) and knots (dot-dot-dashed). The left panel shows the distribution for V-web calculated with a Gaussian smoothing of $R_s=0.25\hmpc$. The right panel shows the distribution after applying the multi-scale correction.
}
\label{fig:hist} 
\end{center} 
\end{figure*}

\begin{table}
\begin{center}
\begin{tabular}{llccc}
web elements & volume filling fraction & mass filling fraction  \tabularnewline
\hline
\hline
voids                &  $0.68\ \ \ (0.69)$          &   $0.13 \ \ \ (0.15)$     \tabularnewline
\hline
sheets               &  $0.27\ \ \ (0.26)$          &   $0.36 \ \ \ (0.37)$     \tabularnewline
\hline
filaments           &  $0.046\ \ \ (0.046)$          &   $0.34 \ \ \ (0.37)$     \tabularnewline
\hline
knots                  &  $0.0036\ \ \ (0.0035)$          &   $0.17 \ \ \ (0.11)$     \tabularnewline
\hline
\end{tabular}
\caption{
 The volume and mass filling factors of the various web elements. The filling factors obtained without the multi-scale correction are given in the parentheses.
}
\label{tab:filling}
\end{center}
\end{table}

\section{Discussion}  
\label{sec:disc}  

A novel algorithm for classifying the cosmic web has been introduced here. The algorithm extends the dynamic classification of Hahn et al (2007) and of FR09 by studying the eigenvalues of the (velocity) shear tensor at any given point rather than those of the (gravitational) tidal tensor. The web classification depends on the number of eigenvalues above a given threshold. Each point is classified as either void (0 eigenvectors above the threshold), sheet (1), filament (2) or a knot (3). The value of the threshold is taken as a free parameter, determined   so as to provide the best visual match to the observed LSS. 
As was pointed out  in the Introduction there is no  model which can make quantitative predictions of the cosmic web at the current epoch non-linear universe. This renders the various cosmic web finders to depend on one or more free parameters that cannot be derived from first principles. The V-web, defined by the threshold and by the adopted resolution, can serve  as a platform for studying, among other things, the properties of galaxies and halos with respect to their environs. The choice of the parameters defining the web needs to be optimized according to the problem at hand.
The focus of the present paper is on the presentation of the method, choosing  the threshold and resolution parameters so as to recover the visual impression that emerges from the LSS of the simulation. Detailed  study and analysis of the cosmic web, its relation to the properties of DM halos and simulated galaxies is to be presented in a forthcoming series of papers. In particular, future studies will focus on studying within the CLUES framework  the Local Group in the context of the local cosmic web. A first step in that direction is  the study  of the orientation of the angular momentum of parent and sub-halos with respect to the cosmic web 
\citep{2012MNRAS.421L.137L}.

Figures \ref{fig:box64-web},  \ref{fig:box8_web_1} and   \ref{fig:box8_web_2}  clearly show  that the kinematic cosmic web (V-web) resolves much smaller structure than the  gravitational cosmic web (T-web).
The two possible definitions of the web coincide on the large scales, where the linear regime prevails, but they depart  with  going to small scales where the non-linear dynamics manifests itself.  Why does the velocity field trace  the underlying cosmic web better? 
The following argument might provide a plausible explanation.
It is  known that the  skeleton of the cosmic web has already been delineated  by the initial conditions \citep{1970A&A.....5...84Z,1996Natur.380..603B}. 
It is also well know that the density field, and hence also the gravity field, evolves away from the linear regime faster than the velocity field 
\citep[][and references therein]{2011arXiv1111.6629K}. 
This  suggests that the velocity field  retains a better memory of the initial conditions than the gravitational field, and therefore is expected to trace  the cosmic web better. The present calculations support that assertion, yet the problem deserves a more in-depth study.

The cosmic web permeates the large scale structure (Bond et al 1996),
covering all regimes from the very under-dense to the densest regions.
Although the cosmic web correlates with the matter density, it cannot
be solely described by the local density. For example, vast
under-dense regions are bisected by planar structures \citep[e.g.][]{2012arXiv1203.0248A} 
that are clearly manifested by the matter
distribution and are tagged as sheets by the V-web finder. The
under-dense blobs, of  $\Delta < 1.0$, are surrounded by two-dimensional
sheets with $\Delta \approx 1.0$. Sheets contain quasi-linear partially collapsed
one-dimensional filaments, which in turn contain highly non-linear
compact knots. Figure 5 shows that the sheets are associated with the
quasi-linear density field. The DM halo mass function has a strong
dependence on the ambient density, and the mass function of halos
residing in low density regions is heavily skewed towards low mass
halos\citep{1999MNRAS.302..111L,2007MNRAS.375..489H}. Hence sheets are
expected to be populated by low mass halos, and therefor to host
predominantly faint galaxies. It follows that magnitude limited
redshift surveys of galaxies which trace the distribution of the more
luminous galaxies are expected to be biased towards the luminous knots
and filaments, leaving behind the dimmer sheets to be virtually
undetected. As telescopes become more and more sensitive and begin to
probe dimmer and dimmer galaxies, the internal structure of under
dense cosmic web types (i.e. voids and sheets) will come into focus.

\section*{Acknowledgments}  

Fruitful discussions with Francisco-Shu Kitaura are gratefully acknowledged.
This research has been partially support by the Deutsche Forschungsgemeinschaft under grant  GO 563/21-1. 
 YH has been partially supported by the ISF (13/08).  
 NIL acknowledges the support of the  DFG and  the hospitality of the KITP, supported in part by the National Science Foundation,  under Grant No. NSF PHY11-25915.
AK is supported by the {\it Spanish Ministerio de Ciencia e Innovaci\'on} (MICINN) in Spain through the Ramon y Cajal programme as well as the grants AYA 2009-13875-C03-02, AYA2009-12792-C03-03, CSD2009-00064, and CAM S2009/ESP-1496.
GY acknowledges support of  MICINN  (Spain) through research grants FPA2009-08958 and  AYA2009-13875-C03-02 and through Consolider-Ingenio  SyeC  (CSD2007-0050).
The simulations were performed at the Leibniz Rechenzentrum Munich (LRZ)
and at Barcelona Supercomputing Center (BSC). We acknowledge the use
of the CLUES data storage system EREBOS at AIP.

\bibliographystyle{mn2e}  
\bibliography{web}  
\end{document}